\documentclass[structabstract]{aa}  

\usepackage{graphicx}
\usepackage{txfonts}
\usepackage{longtable}
\usepackage{lscape}
\usepackage{natbib}
\bibpunct{(}{)}{;}{a}{}{,} 

\begin{document}

   \title{The relationship between gas content and star formation rate in spiral galaxies.
   Comparing the local field with the Virgo cluster.}

   \subtitle{}

   \author{Michele Fumagalli
          \inst{1,2}
          \and
          Giuseppe Gavazzi\inst{1}
          }

   \institute{Istituto di Fisica G. Occhialini, Universita' di Milano-Bicocca,
              Piazza della scienza 3. Milano - Italy\\
              \email{giuseppe.gavazzi@mib.infn.it}
         \and
             Department of Astronomy and Astrophysics, University of California, Santa Cruz, CA 95064\\
            \email{mfumagal@ucsc.edu}
             }

   \date{}

  \abstract
   {Despite many studies of star formation in spiral galaxies, a complete and coherent
understanding of the physical processes that regulate the birth of stars has not yet been 
achieved, nor has unanimous consensus been reached, despite the many attempts, 
on the effects of the environment on the star formation in galaxy members of rich clusters.}
   {We focus on the local and global Schmidt law and we investigate how cluster galaxies have their star formation activity perturbed.}
   {We collect multifrequency imaging for a sample of spiral galaxies, members of the Virgo cluster and of the local field; we compute the surface density profiles for the young and for the bulk of the stellar components,  for the
molecular and for the atomic gas.}
   {Our analysis shows that the bulk of the star formation correlates with the molecular gas, but the atomic gas is important or even crucial in supporting the star formation activity in the outer part of the disks.
Moreover, we show that cluster members that suffer from a moderate HI removal have their molecular component and their SFR quenched, while highly perturbed galaxies show an additional truncation in their star forming disks.}
   {Our results are consistent with a model in which the atomic hydrogen is the fundamental fuel
for the star formation, either directly or indirectly through the molecular phase; therefore galaxies whose
HI reservoirs have been depleted suffer from starvation or even from truncation of their star formation
activity.}

   \keywords{Galaxies: evolution -- galaxies: ISM -- galaxies: spiral -- galaxies: clusters: individual (Virgo) -- stars: formation}
   
   \authorrunning{Fumagalli \& Gavazzi}
   \titlerunning{Gas content and star formation rate in spiral galaxies.}
   \maketitle
%

\section{Introduction}

A satisfactory understanding of the physical processes governing the formation of giant molecular clouds (GMC) from 
primeval atomic hydrogen (HI) and their instability leading to the formation of stars is still far from achieved.
This is not surprising due to the complex mix of hydrodynamical and gravitational 
physics involved in the process of star formation.
Beside the theoretical difficulties, what surprisingly limits our present understanding of this issue
is the lack of a robust phenomenology of star formation in nearby galaxies. 
Despite the many recipes that have been proposed\footnote{See for instance the dynamically modified Schmidt law
in \citet{won02}}, the simplest and most commonly used parametrization is 
the Schmidt law \citep{sch59} that establishes a correlation between the star formation 
rate density $\rho_{sfr}$ and the gas density $\rho_{gas}$:
\begin{equation}
\rho_{sfr}=a\rho_{gas}^\alpha\:;
\end{equation}
derived for our Galaxy, the validity of this law has been tested also for external galaxies \citep[e.g.][]{ken83},
for which a similar relation, usually written in terms of surface densities, holds:
\begin{equation}\label{Slaw}
\Sigma_{sfr}=A\Sigma_{gas}^n\:.
\end{equation}
Many studies of the global Schmidt law, i.e. the relationship between the disk--averaged
star formation rate (SFR) and the total gas content, have been carried out on large samples of galaxies.
A milestone paper on this issue is by \citet{ken98}, who studied the connection between the
integrated SFR surface density and the integrated gas surface density in a sample of normal and
starburst galaxies. 
Using integrated quantities, he found that the same Schmidt law with an index $n=1.4$ holds for 
both normal and starburst galaxies, covering a range of 5 dex in the gas density and over 6 dex 
in the SFR; he also found that this relation is mainly driven by the correlation between the SFR 
and the HI, while there is a poorer correlation between the SFR and the H$_2$, 
significant only when massive (L$_B>10^{10}$ L$_\odot$) objects are considered. 
Owing to the fact that star formation takes place in molecular regions on parsec scales, this result was 
somewhat unexpected, but Kennicutt argued that the poor knowledge of the CO--to--H$_2$ 
conversion factor might be responsible for the scatter in the H$_2$/SFR correlation. 
Using seven galaxies from the BIMA survey \citep{hel03}, \citet{won02} studied the applicability 
of the Schmidt law on local scales, i.e.
through a comparison between the SFR and the gas surface density profiles. 
They found  that a Schmidt law between the SFR and the gas content holds also on local scales and, 
even if the molecular gas content alone does not always control the SFR, they concluded that the 
correlation found for the total gas is entirely driven by the molecular component.    
Consistent results were found by \citet{boi03}, who used low resolution CO data but included 
an $X$ factor that varies as a function of the metallicity.  
Another open issue is to asses the influence of the environment on the star formation.
Galaxies in clusters suffer from a variety of environmental perturbations, mainly due to 
gravitational interactions between  galaxies themselves or between a galaxy and the cluster potential well
\citep[see the review by][]{bos06} or to various hydrodynamical interactions between the ISM and the intergalactic medium (IGM).
Tidal interactions between galaxies or between individual galaxies and the cluster potential \citep{mer84,
byr90} produce both the 
removal of the outer and looser components and the gas infall towards the galaxy center; 
harassment \citep{moo96} can induce sinking of gas towards the galaxy center and can shape 
the stellar profiles; the ram pressure stripping \citep{gun72} or the viscous stripping \citep{nul82} 
can effectively remove the gas components, mainly from the outer part of the disks.
Furthermore, a combination of gravitational and hydrodynamic effects can cause galaxy
starvation or strangulation \citep{lar80,bek02}. 
According to starvation, the gas that feeds the star formation in the local universe 
probably comes from the infall of an extended gas reservoir, therefore the effect of removing the outer 
galaxy halo would be that of preventing further gas infall; as a consequence,  the star formation 
exhausts the available gas, quenching further activity.
The literature about the classification and the effects produced by the environment on the evolution of cluster galaxies 
is very rich \citep[e.g.][]{bos06}, but, despite many studies on nearby and distant clusters,
it is not clear what the dominant processes are, nor how the star formation
activity is affected \citep[and reference therein]{koo04}. Unanimous consent seems only to hold
on the statement that environmental processes are effective at removing the outer ISM 
and therefore affecting the HI component \citep{cay94,gio85}, leaving the molecular component, 
well--bounded inside the galaxy potential well, undisturbed \citep{ken89, bos02c}.
In order to study the environmental effects on star formation, we collect both objects that are
members of the Virgo cluster and local field galaxies. 
Since the different components involved
in the star formation are observable along a broad stretch of the electromagnetic spectrum, 
we adopt a multifrequency analysis. 
To quantify the atomic hydrogen we collect observations at 21 cm;
the molecular hydrogen content is estimated indirectly via CO 
observations at 2 mm; we quantify the stellar mass  
with observations at the near--infrared (NIR) or visible bands and 
we study the presence of new stars indirectly, observing the hydrogen recombination line (H$\alpha$)
at 6563 \AA.
The sample is presented in Sec. \ref{data}, together with the description of the data reduction
procedures; the analysis is given in Sec. \ref{analysis}, while we discuss our results and conclusions
in Sec. \ref{results} and \ref{concl}.    
  
\section{Data}\label{data}
\subsection{The sample}

The sample of galaxies chosen for the present analysis is selected primarily according to the availability of 
high sensitivity, extensive CO mapping, such as in  
the {\it Nobeyama CO atlas of nearby spiral galaxies} by \citet{kun07}. 
Secondly we require that reliable H$\alpha$ imaging exists in the literature. 28 giant spiral galaxies match these criteria.  For all of them we were able to find additional imaging material in the red continuum ($i$ or H band) and for 25/28 we found sensitive HI mapping.
Table \ref{sample} lists some observational parameters and references for the selected galaxies.
The sample is well balanced between isolated objects and members of the Virgo cluster, as stressed by the local density parameter (see Col. 11).
One characteristic of our sample is that it is composed
of giant spirals with a typical H--band luminosity of $10^{10}-10^{11}$ L$_\odot$. No BCDs nor Irr/dIrr are included.
The fact that galaxies in the Nobeyama atlas have not been
selected according to a CO emission criterion \citep[Sec. 2]{kun07} protects against a possible
bias towards CO rich galaxies. 
One could question why we preferred the Nobeyama atlas to higher resolution interferometric CO material, such as
the {\it BIMA Survey of Nearby Galaxies} \citep{hel03} that 
provides images for 44 nearby spiral galaxies.
This is mainly to ensure that our flux measurements are unaffected by severe missing flux problems 
on scales larger than $20-30''$ at 115 GHz. Even though spectra from the central galaxy regions
have been used to recover the  missing flux or single 
dish observations provide the zero-spacing data, the algorithms used to recover the extended flux are
somewhat unreliable \citep[see][App. A]{hel03}. 
 
\begin{table*}
\centering
\begin{tabular}[h]{c c c c r r c c c c c c c c c}
\hline
\hline
NGC &   Name  &RA       & DEC       & cz   & Dist.   &Incl.&	Type&   H lum.   &  r$_{25}$   & $\rho$ &  H$\alpha$ & Met. & HI & Opt.\\
\hline
   &      & hhmmss       & ddppss      & km/s   & Mpc   &deg.&	 &      L$_\odot$   &  arcsec   &  Mpc$^{-3}$ &   & &   &  \\
 (1)   &  (2)     &  (3)    &  (4)      & (5)	& (6)  & (7) & (8)   & (9)         &(10)   & (11)   & (12) & (13) & (14)& (15) \\
\hline
 1068  &  U2188 & 024240.7&  -000048  & 1137 & 14.4 & 21 & SSb   &  11.14 & 185 & 0.34 &  d  &  l  &  -  &  r \\
 3184  &  U5557 & 101817.0&  +412528  &  593 & 8.7  & 24 & SABc  &  10.24 & 222 & 0.17 &  e  &  l  &  n  &  r \\
 3351  &  U5850 & 104357.7&  +114213  &  778 & 8.1  & 41 & Sb	 &  10.36 & 217 & 0.54 &  g  &  m  &  n  &  r \\
 3521  &  U6150 & 110548.5&  -000209  &  802 & 7.2  & 65 & SABb  &  10.63 & 249 & 0.19 &  f  &  m  &  n  &  r \\
 3627  &  U6346 & 112015.0&  +125929  &  721 & 6.6  & 57 & SABb  &  10.51 & 307 & 0.35 &  g  &  k  &  n  &  r \\
 4051  &  U7030 & 120309.6&  +443152  &  704 & 17   & 30 & SABb  &  10.57 & 147 & 1.06 &  e  &  -  &  q  &  r \\
 4102  &  U7096 & 120623.1&  +524239  &  845 & 17   & 58 & SABb  &  10.55 &  89 & 1.42 &  h  &  -  &  q  &  r \\
 4192  &  V0092	& 121348.3&  +145401  &  195 & 17   & 78 & Sb	 &  10.99 & 329 & 1.44 &  a  &  -  &  p  &  s \\
 4212  &  V0157 & 121539.3&  +135405  &  -85 & 17   & 49 & Sc	 &  10.34 &  84 & 1.44 &  b  &  -  &  -  &  r \\
 4254  &  V0307 & 121849.6&  +142459  & 2733 & 17   & 24 & Sc	 &  10.94 & 150 & 1.91 &  a  &  i  &  p  &  s \\
 4303  &  V0508 & 122154.9&  +042825  & 1912 & 17   & 35 & Sc	 &  10.98 & 207 & 1.06 &  a  &  i  &  p  &  s \\
 4321  &  V0596 & 122254.9&  +154921  & 1898 & 17   & 27 & Sc	 &  11.14 & 181 & 2.95 &  a  &  i  &  o  &  s \\
 4402  &  V0873 & 122607.5&  +130646  &  562 & 17   & 74 & Sc	 &  10.39 & 106 &   -  &  c  &  j  &  p  &  s \\
 4419  &  V0958 & 122656.4&  +150251  &   50 & 17   & 70 & Sa	 &  10.61 & 117 & 1.60 &  a  &  -  &  o  &  s \\
 4501  &  V1401 & 123159.2&  +142514  & 2606 & 17   & 58 & Sbc   &  11.18 & 189 & 2.04 &  a  &  i  &  o  &  s \\
 4535  &  V1555 & 123420.3&  +081152  & 2294 & 17   & 27 & Sc	 &  10.76 & 244 & 2.33 &  a  &  j  &  p  &  s \\
 4536  &  V1562 & 123427.1&  +021116  & 2147 & 17   & 63 & Sc	 &  10.71 & 212 & 0.43 &  a  &  j  &  o  &  s \\
 4548  &  V1615 & 123526.4&  +142947  &  804 & 17   & 34 & Sb	 &  10.91 & 165 & 2.82 &  a  &  -  &  p  &  s \\
 4569  &  V1690 & 123649.8&  +130946  &  106 & 17   & 62 & Sab   &  11.05 & 274 & 1.15 &  a  &  k  &  o  &  s \\
 4579  &  V1727 & 123743.5&  +114905  & 1844 & 17   & 40 & Sab   &  11.11 & 150 & 3.26 &  a  &  k  &  p  &  s \\
 4654  &  V1987 & 124356.6&  +130735  & 1357 & 17   & 59 & Sc	 &  10.66 & 140 & 2.93 &  a  &  i  &  o  &  s \\
 4689  &  V2058 & 124745.5&  +134546  & 1934 & 17   & 41 & Sc	 &  10.48 & 114 & 2.39 &  a  &  i  &  o  &  s \\
 4736  &  U7996 & 125053.0&  +410713  &  314 & 4.3  & 35 & Sab   &  10.45 & 233 & 0.42 &  e  &  m  &  n  &  r \\  
 5055  &  U8334 & 131549.3&  +420145  &  508 & 7.2  & 56 & Sbc   &  10.71 & 352 & 0.40 &  g  &  m  &  n  &  r \\  
 5194  &  U8493 & 132952.7&  +471142  &  461 & 7.7  & 30 & Sbc   &  10.84 & 233 & 0.33 &  g  &  m  &  n  &  r \\  
 5236  &  UA366 & 133700.9&  -295155  &  508 & 4.7  & 46 & Sc	 &  10.72 & 465 & 0.18 &  f  &  m  &  n  &  t \\  
 5457  &  U8981 & 140312.5&  +542056  &  237 & 5.4  & 22 & SABc  &  10.47 & 720 & 0.29 &  e  &  l  &  n  &  r \\  
 6951  &  U11604& 203714.0&  +660620  & 1424 & 24.1 & 52 & SABb  &  11.10 &  97 & 0.08 &  h  &  -  &  -  &  u \\  
\hline														     
\end{tabular}
\caption{Observational parameters. 
Column 1,2: NGC, VCC or UGC designations. 
Column 3,4: coordinates (J2000) from GOLDMine and HyperLeda. 
Column 5: recessional velocities in km/s from GOLDMine and HyperLeda.
Column 6: distances in Mpc, assigned following the subcluster membership criteria of \citet{gav99} or from the 
{\it Nearby Galaxies Catalogue} by \citet{tul94}. 
Column 7: galaxy inclinations in deg from GOLDMine and HyperLeda.
Column 8: morphological types from GOLDMine and HyperLeda. 
Column 9: H--band luminosity in L$_\odot$ from GOLDMine or computed from H magnitude from SIMBAD.
Column 10: radius in arcsec at the B--band 25th isophote from  HyperLeda. 
Column 11: density in Mpc$^{-3}$ for galaxies brighter than -16 mag in the vicinity of the galaxy considered, determined on a 3D--grid at 0.5 Mpc spacing 
\citep{tul94}. 
Column 12: reference to H$\alpha$ imaging: a = \citet{bos02}; b = \citet{koo01}; c = \citet{bos02b}; d = \citet{mar01}; e = \citet{kna04}; 
f = \citet{meu06}; g = \citet{ken03}; h = \citet{jam04}. 
Column 13: reference to metallicity: i = \citet{ski96}; j = GOLDMine; k = \citet{ken03}; l = \citet{van98}; m = \citet{zar94}. 
Column 14: reference to HI mapping: n = THINGS; o = VIVA; p = \citet{cay90};  q = \citet{ver01}. 
Column 15: reference to NIR/Visual imaging: r = SDSS; s = GOLDMine; t = \citet{kuc00}; u = \citet{mul97}.}\label{sample}
\end{table*}

\subsection{H$\alpha$ data}
We compute the star formation rate from the intensity of the H$\alpha$ emission line, following the method of  
\citet{ken83,ken98}. H$\alpha$ images for the VCC galaxies are retrieved from GOLDMine \citep{goldmine}, while
for the remaining galaxies we retrieved images from NED\footnote{See table \ref{sample} 
for individual references.}.
Images from GOLDMine are calibrated following \citet{bos02} and \citet{bos02b}, i.e. using
the photometric zero-point included in the image headers. For the images collected from NED instead, we derive
the photometric calibration using published fluxes \citep{you96,rom90,ken83d}\footnote{We correct fluxes from \citet{ken83d} with the factor 1.16, as suggested by the authors}.
Before converting the H$\alpha$ flux into a star formation rate, we correct for the [NII] emission;
for almost all the galaxies we apply a correction derived from drift-scan spectra that accounts
for the mean [NII] emission from the entire disk. Where individual correction factors are not
available, we apply the standard value $\rm [NII]/H\alpha = 0.53$  \citep{ken92}. 
Since we consider [NII]/H$\alpha$ factors averaged over the entire disk, the correction underestimates
the [NII] emission from the bulges of galaxies that host active galactic nuclei (AGN), where nuclear [NII] can be even stronger than H$\alpha$. This is particularly severe in our sample composed of giant spirals, 
since almost all massive galaxies are known to host an AGN \citep{dec07}. In addition, even if the nuclear emission was corrected by the suitable factor, 
it would not be possible to disentangle the H$\alpha$ flux due to star formation from the one due to the AGN ionization. 
Therefore, by inspecting both the H$\alpha$ imaging and the available spectra, we identify a nuclear region for each galaxy, and in the following analysis we do not include the emission from these regions.
As for the dust absorption, since the H$\alpha$/H$\beta$ or the  H$\alpha$/Br$_\gamma$ ratios
are unknown for the majority of our galaxies, we prefer to apply a standard dust extinction $A=1.1$ mag \citep{ken83} to other more sophisticated, yet unreliable corrections.
From the corrected H$\alpha$ luminosity we compute the star formation rate using the conversion \citep{ken98}
\begin{equation}\label{calsfr}
{\rm SFR\: [M_\odot yr^{-1}]=7.94 \times 10^{-42}\: L(H\alpha)\:[erg/s]}\:,
\end{equation}
based on a Salpeter IMF ($\alpha=2.35$) over $\rm 0.1-100M_{\odot}$.
Furthermore we assume that a fraction $f$=0.57 of Lyman continuum photons ionizes the hydrogen atoms inside the star forming regions, where dust is mixed with gas \citep{hir03}.
On the contrary, we neglect the correction for the escape fraction of continuum Lyman photons, since it
has been shown that it is less than 6\% in starburst galaxies and even negligible for normal galaxies \citep{hec01}.
We estimate that the typical flux uncertainty is 0.05 mag for VCC galaxies;  0.1 mag
for fluxes from \citet{ken83d},  5\% for those from \citet{you96} and up to 20\% for \citet{rom90}. 
These errors produce a final uncertainty on the star formation rate which ranges from 0.04 dex to
0.09 dex, with a mean value of 0.05 dex. An additional uncertainty of 0.2 dex affects all data due to the uncertainty in the dust correction.
\begin{figure*}
\centering
\begin{tabular}{c c}
\includegraphics[width=8cm]{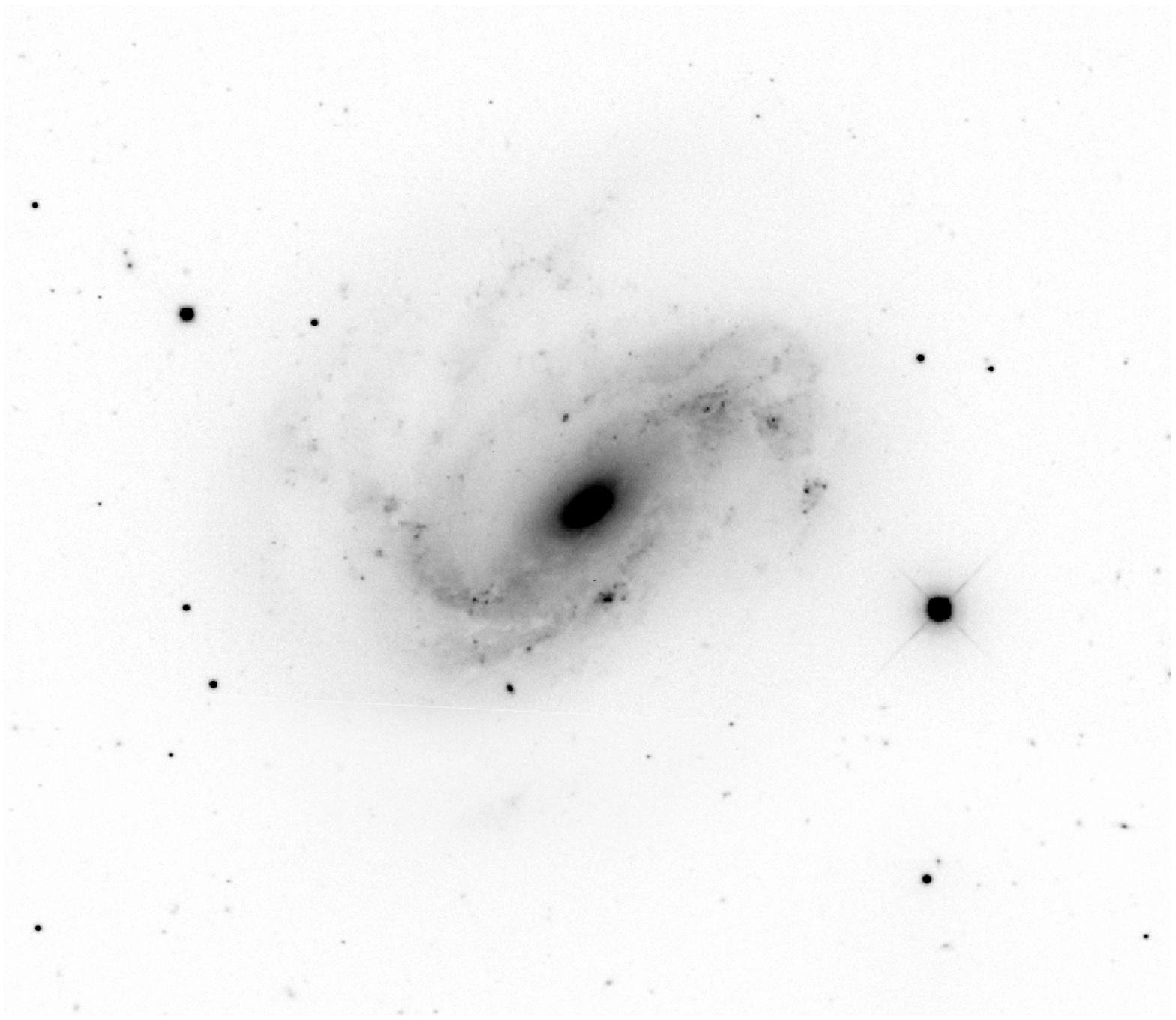}&\includegraphics[width=8cm]{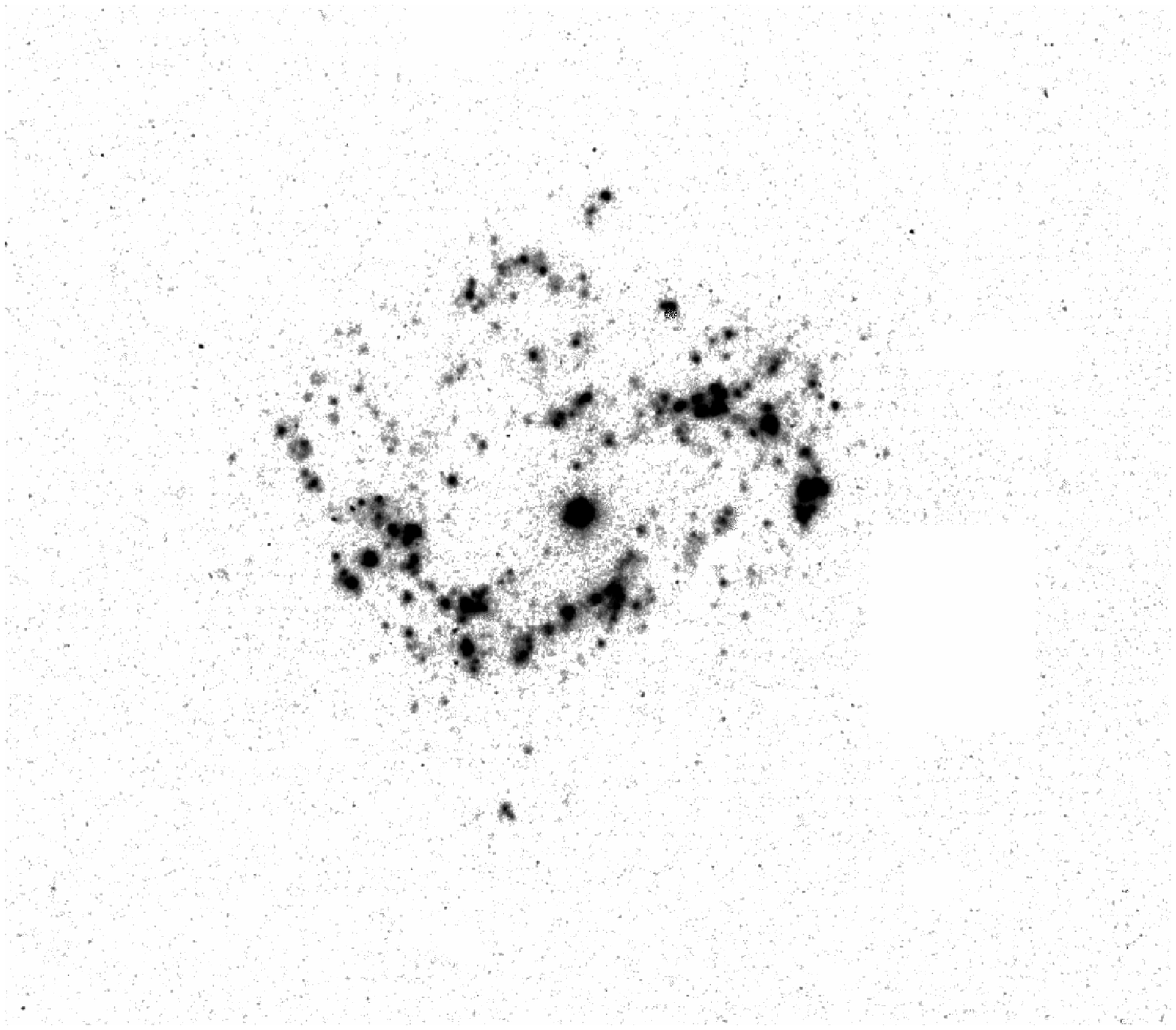} \\
\includegraphics[width=8cm]{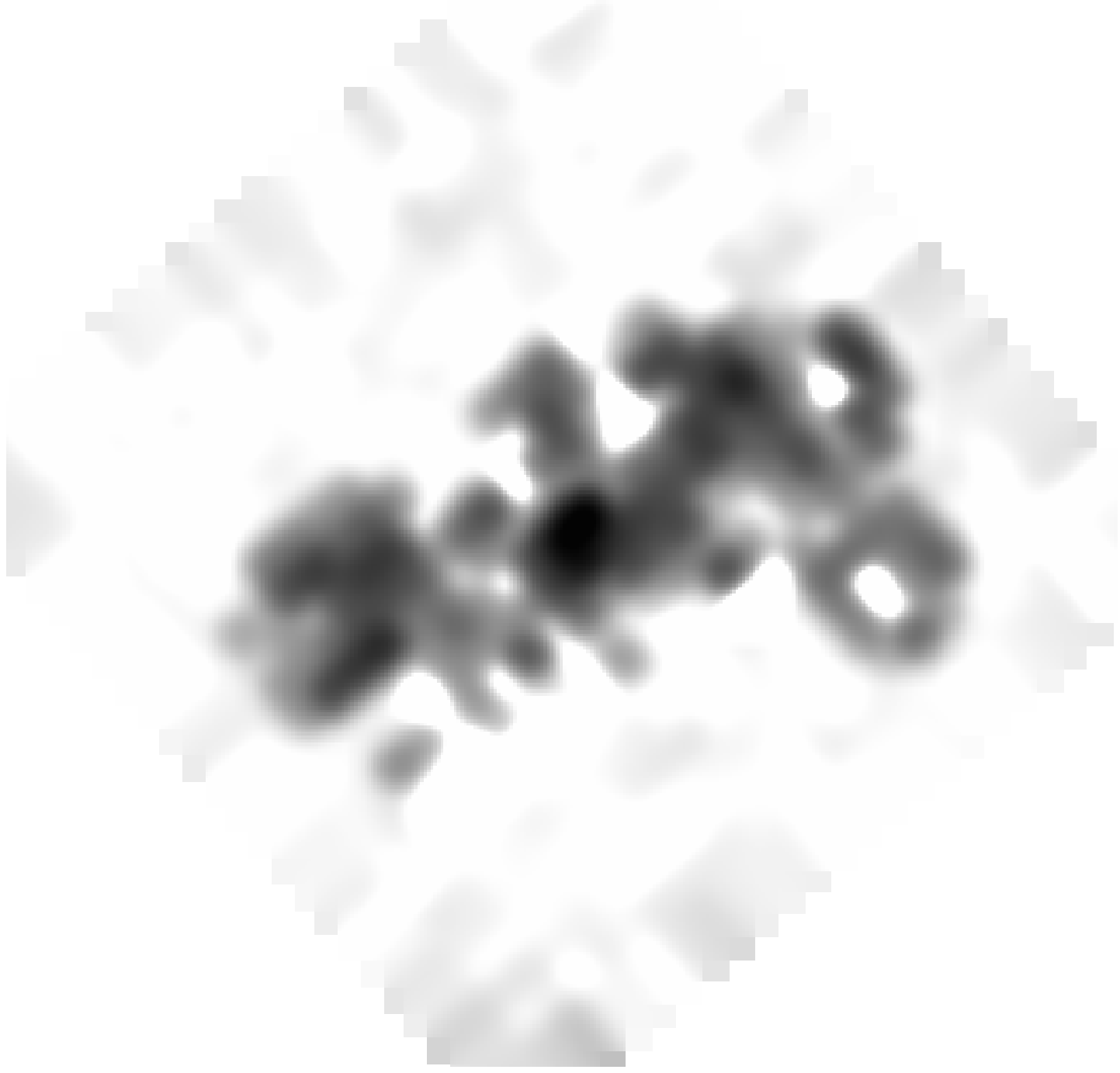}&\includegraphics[width=8cm]{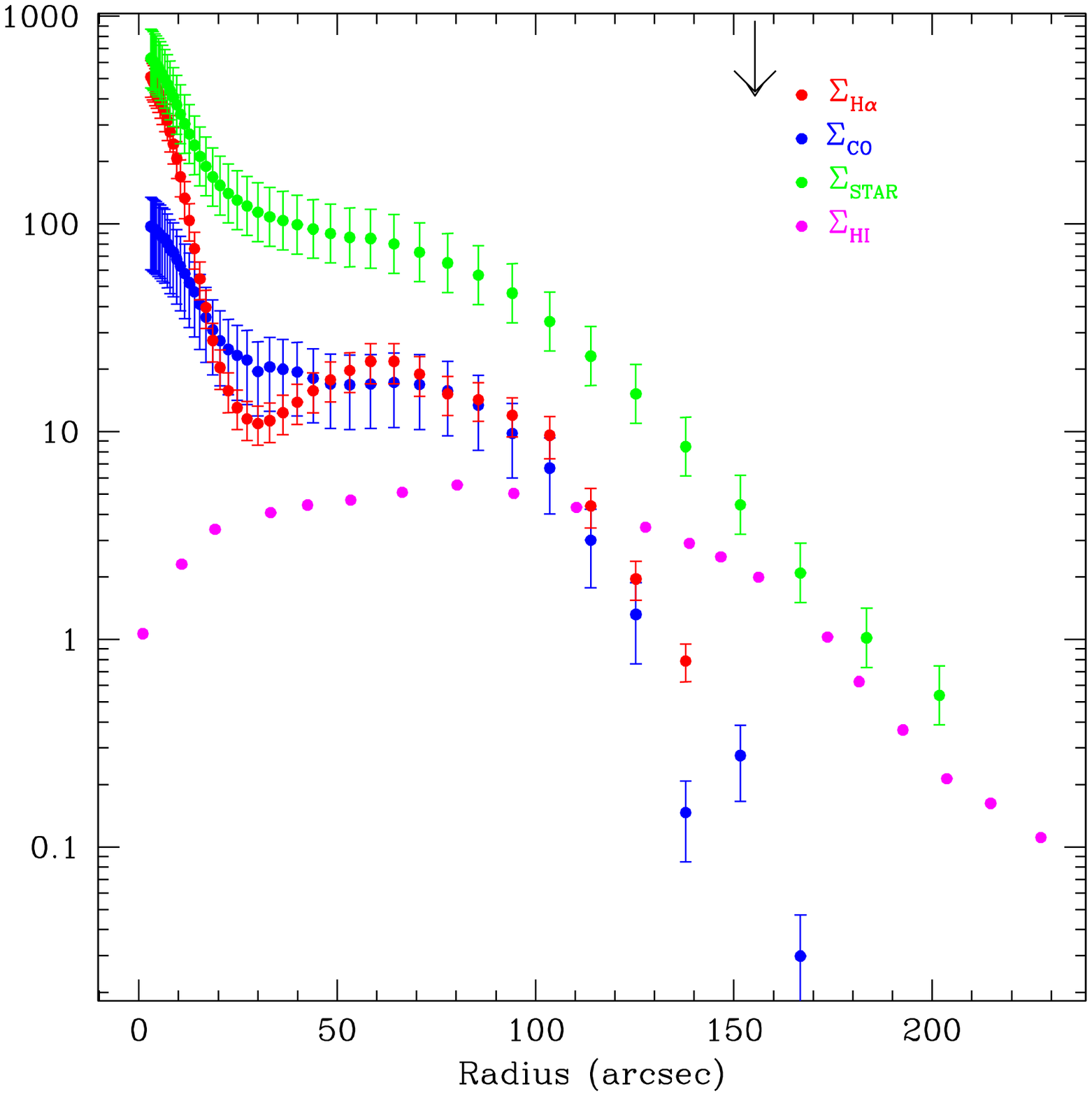}\\
\end{tabular}
\caption[ ]{Example of multiband images and profiles: NGC 4051. Top--left: $i$ band;
top--right: H$\alpha$; bottom-left: CO; bottom-right: surface brightness profiles.
$\Sigma_{sfr}$ (red) in M$_\odot$ pc$^{-2}$ Gyr$^{-1}$; $\Sigma_{HI}$ (magenta), $\Sigma_{H_2}$ (blue) and $\Sigma_{star}$ (green) in M$_\odot$ pc$^{-2}$. The arrow indicates $r_{25}$ as in Col. 10, Tab. \ref{sample}.}
\label{4051}
\end{figure*}

\subsection{H$_2$ data}
In the absence of a direct tracer of the H$_2$ molecule, the  molecular
hydrogen abundance can be inferred from CO observations. 
We collect flux calibrated maps of the $\rm ^{12}CO(J:1-0)$ transition at 115 GHz from the {\it Nobeyama CO atlas of nearby spiral galaxies} taken with the 45--m radiotelescope by \citet{kun07}.
Despite their spatial resolution of 15 arcsec (at 115 GHz), these observations have been preferred
to higher resolution interferometric data because of their higher sensitivity to the large--scale emission from galaxy disks. In fact the typical rms noise of 40--100 mK combined with a velocity resolution of 5 km/s, gives a sensitivity of $\rm N(H_2)=5-10 \times 10^{19}\:cm^{-1}$.
The most challenging task related with the CO observations, both from a theoretical and an observational point of view, is the determination of the CO--to--H$_2$ conversion factor ($X$ in $\rm cm^{-2}/(K\:km/s)$). 
Many attempts to determine a reliable CO--to--H$_2$ conversion factor can be found in the literature;
nevertheless, there is not yet unanimous consent on the dependence of $X$ on parameters such as density, temperature or even the geometry of the clumps that constitute the molecular regions \citep{wal07}. 
Due to these difficulties, a constant conversion factor $X$, as derived for our Galaxy,
has been often assumed, probably underestimating the H$_2$ in low luminosity galaxies.
In addition, it is expected that the CO--to--H$_2$ conversion strongly depends on dust and metallicity that 
regulate the H$_2$ formation and photodissociation or the CO heating.
Therefore, we avoid a single conversion factor for our sample, but we choose individual $X$ as a function of the metallicity, following the empirical calibration by \citet{bos02c}, made on fourteen independent measurements 
of CO and molecular hydrogen. 
\begin{equation}\label{xmet}
\rm \log X=-1.01\pm0.14 \times \left(12+\log(O/H)\right) + 29.28\pm0.20\:.
\end{equation} 
With this equation we compute the conversion factor for each galaxies (See Col. 13 in Table \ref{sample}
for references to metallicity data). 
For seven galaxies without published 12+log(O/H) we derive the metallicity using \citep{bos02c}
\begin{equation}\label{extrmet}
\rm 12+\log(O/H)=0.37\pm0.04 \times \log L_H\:[L_\odot]+4.98\pm 0.15\:,
\end{equation} 
calibrated on a sample of nearby and Virgo cluster galaxies. 
The mean discrepancy between the metallicity directly measured from spectra and derived using (\ref{extrmet}) is 
0.2 dex in our sample, not much greater than the uncertainties on the measurements themselves and significantly lower
than the statistical uncertainty on the X conversion factor derived using (\ref{xmet}), that is on average 1.3 dex. One should keep in mind that other unknown dependencies, beside the one on metallicity, would increase
the uncertainty on $X$. 
Since spiral galaxies exhibit a radial metallicity gradient, when available in the literature, we account also for them to compute $X$ as a function of the galaxy radius. We find that $X$ spans 0.4 dex on average in individual galaxies and in some cases up to 1 dex. 
With the $X$ conversion factor, we compute the molecular hydrogen column density using:
\begin{equation}
\rm N(H_2)\:[M_\odot\:cm^{-2}]=2\times1.36\times m_p\times X \times I_{co}\:,
\end{equation}	   
where 1.36 is the helium correction factor, $m_p$ is the proton mass and I$_{co}$ is the CO flux in K km/s.

\subsection{HI data}
To study the distribution of the HI mass, we collect images at 21 cm from various sources\footnote{See Table \ref{sample} for individual references.}.
For seven Virgo galaxies, we obtained moment 0 maps from VIVA, {\it The VLA Imaging of Virgo in Atomic gas} \citep[Chung et al. 2008, in preparation;][]{chu07}; VIVA images have a 
spatial 
resolution of $15-17$ arcsec and 
they reach a
typical sensitivity of $1.5\times 10^{19}$ cm$^{-2}$.
For nine field galaxies, instead, we obtain HI images from THINGS, {\it The HI Nearby Galaxy Survey} 
\citep{wal08}; THINGS images have a spatial resolution of 11 arcsec and a sensitivity
of $1.2\times 10^{19}$ cm$^{-2}$. For the remaining galaxies, we collect HI data from the literature,
either from \citet{cay90} or from \citet{ver01}. These maps have a lower spatial resolution, spanning
from 40 to 70 arcsec. The sensitivity ranges from $1\times 10^{19}$ cm$^{-2}$ to $1\times 10^{20}$ cm$^{-2}$.
For three galaxies (NGC 1068, NGC 4212 and NGC 6951) we cannot find HI maps, but only a measurement of their
integrated HI mass.
From the flux calibrated images, we compute the neutral hydrogen surface density using
\begin{equation}
\rm N(HI)\:[M_\odot\:cm^{-2}]=1.36\times m_p\times I_{HI}\:,
\end{equation}	   
where  1.36  is the correction factor for the helium and I$_{HI}$ is the column density in cm$^{-2}$.
The total HI mass in solar units is computed using:
\begin{equation}
\rm \log M_{HI}\:[M_\odot]=\log(2.36\times10^5\times SintP\times D^2)\:,
\end{equation}
where the total HI flux SintP (Jy km/s) is from ALFALFA, {\it The Arecibo Legacy Fast ALFA Survey} \citep{gio05}, or from the HyperLeda database and D is the galaxy distance in Mpc.

\subsection{Optical data}
The stellar mass distribution is obtained from imaging in the NIR H band from GOLDMine or the $i$ band
from the {\it Sloan Digital Sky Survey}  \citep[SDSS-DR6,][]{ade08}. Exceptions are 
NGC 5236 and NGC 6951 (not imaged in the SDSS) for which we collect respectively I band and K band images from NED. In Table \ref{sample} we show individual references.
Due to their large angular dimension, our galaxies are often cut in two or more SDSS fields, thus we retrieve images using $Montage$, a toolkit for assembling FITS images over multiple frames into a mosaic.
The photometric calibration for galaxies from GOLDMine relies on the zero-points provided in the image
headers, obtained according to \citet{gav96}; GOLDMine images
for NGC 4192 and NGC 4569 are from  {\it The Two Micron All Sky Survey}  \citep{skr06}, thus for these galaxies we adopt the 2MASS standard calibration procedure. For all images, the typical uncertainty is $<$0.05 mag.
SDSS images are calibrated (within 1\% accuracy) using the zero-points derived from a set of parameters (calibration constant, air mass and extinction coefficient) provided in the tables associated with the images. 
To derive the stellar mass, we apply to each galaxy a mass--to--light ratio, computed through an extrapolation along the IMF according to \citet{bel03}:
\begin{equation}\label{mass2lig}
\rm \log (M/L)_\lambda\:[M_\odot/L_\odot]=a_\lambda+b_\lambda \cdot Color\:,
\end{equation}
with $a_\lambda$ and $b_\lambda$ chosen accordingly to the band used. 
We use B--V $Color$ from HyperLeda in combination with NIR data and 
$g$--$i$ for $i$ band data. The mean dispersion on M/L is 0.1 dex.

\subsection{The surface brightness profiles}
For each galaxy we derive a calibrated surface brightness profile at four frequencies.
All images are firstly astrometrically calibrated using GAIA, then aligned and resampled to 3 arcsec/pixel.
Stars near the galaxies are masked using the IRAF task {\tt imedit} and the local sky background is estimated and subtracted using {\tt marksky} and {\tt skyfit} from Galphot\footnote{GALPHOT
was developed for IRAF-STSDAS by W. Freudling, J. Salzer, and
M.P. Haynes and adapted by S.Zibetti and L. Cortese. See \citet{gav01}.}.
Using  a Gaussian kernel we smooth all images to a spatial resolution of 15 arcsec, in common with the CO observations and most of the HI images. 
To derive the surface brightness profiles we use the IRAF task {\tt ellipse}
that fits elliptical isophotes to the galaxy images.
Starting from a set of initial parameters given manually for the NIR or $i$ band
imaging, the fit maintains as 
free parameters the ellipse center, the ellipticity and the position angle;
the ellipse semi--major axis is incremented by a fixed fraction of its value at each step of the fitting procedure and the routine halts when the surface brightness found in a given annulus equals twice the sky rms.
This task provides the statistical error of the surface brightness.
From the same set of ellipses we derive the surface brightness profiles for the remaining images,
with the exception of the HI data from \citet{cay94} and \citet{ver01} for which we interpolate the published 
profiles on our grid.
The derived surface density profiles are corrected for projection effects using the galaxy inclination
on the plane of the sky.
For each galaxy we finally obtain  the HI, H$_2$ and stellar surface density profiles 
$\Sigma_{HI}$, $\Sigma_{H_2}$ and $\Sigma_{star}$ in $\rm M_\odot\:pc^{-2}$; the star formation rate surface density profiles $\Sigma_{sfr}$  are in $\rm M_\odot\:pc^{-2}\:Gyr^{-1}$.
Moreover, we compute the color profiles: $sfr-sta=\log \Sigma_{sfr} - \log \Sigma_{star}$,~
$sfr-H_2=\log \Sigma_{sfr} - \log \Sigma_{H_2}$ and $H_2-sta=\log \Sigma_{H_2} - \log \Sigma_{star}$. 
Fig. \ref{4051} illustrates an example of the imaging material and density profiles for the galaxy
NGC 4051. From top-right (anticlockwise): H$\alpha$ net; $i$-band (SDSS); CO image; surface brightness profiles 
of stars (green), HI (magenta), SFR (red) and H$_2$ (blue).

\section{Analysis}\label{analysis}

\subsection{The Schmidt law}

Following \citet{ken98}, \citet{won02} and \citet{boi03} we test the validity of the Schmidt law on both local and global scales.

\subsubsection{The local Schmidt law}

\begin{figure*}[t]
\centering
\begin{tabular}{c c}
\includegraphics[width=8cm]{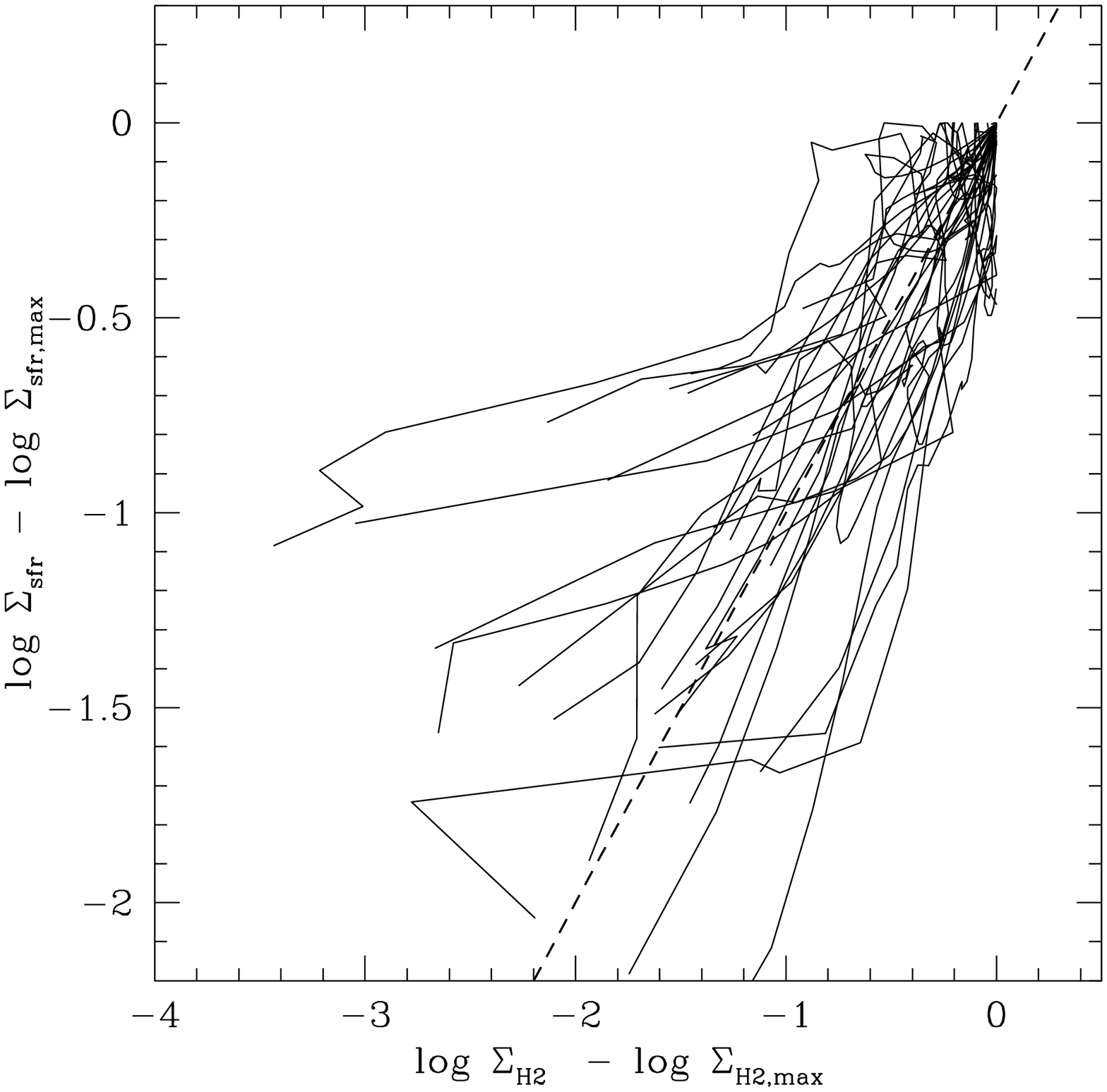}&\includegraphics[width=8cm]{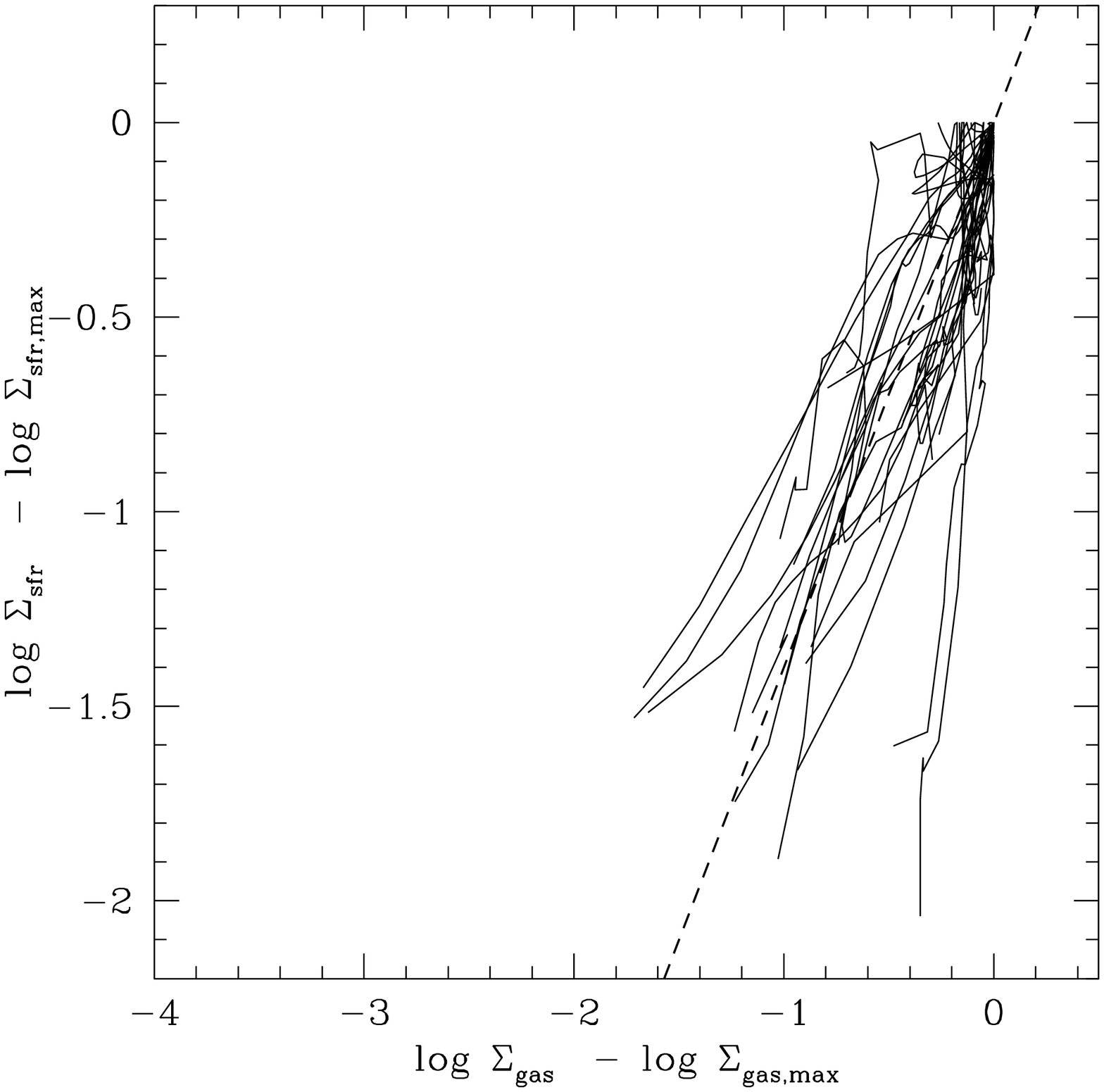}\\
\end{tabular}
\caption{The local Schmidt law for the entire sample, separately for the molecular (left panel) and the total gas (right panel). Profiles are normalized to the maximum values for each galaxy. The dashed line in the left panel represents the index $n=1$, in the right panel $n=1.4$.}\label{schloch}
\end{figure*}

Figure \ref{schloch} (left) shows the correlation between the SFR 
and the H$_2$ surface brightness in the individual galaxies belonging to the studied sample; in doing this comparison, we exclude
the nuclear regions as previously discussed. Since in each galaxy the profiles do not have exactly the same radial extent, we prefer to truncate the most extended profile at the radius of the shortest one, avoiding any extrapolation. Typically, the SFR and the H$_2$ emission have a comparable spatial extent, in which
on average there is a fair correlation between the SFR and the molecular
hydrogen surface density. In fact 80\% of our sample shows
a correlation index $r>0.8$. Only a few galaxies show a weak correlation, mainly due to peculiarities such as prominent bars empty of star formation (NGC 4303), starburst rings (NGC 4321) or significant gas displacements due to known severe environmental perturbations (NGC 4548).
For the correlated galaxies, we perform a least square fit; despite the overall validity of the Schmidt law, 8/20 objects cannot be fitted using a single power law or the obtained rms is greater than 0.15 dex, arbitrarily assumed as a threshold. 
Inspection of the left panel of Figure \ref{schloch} shows that the faintest points in the SFR/H$_2$ relation that belong to the outer parts of the disks significantly deviate from the correlation driven by the points associated with the inner part of the galaxy disks. For the remaining 60\% of the correlated galaxies, we obtain a reliable fit using a single power law;  the mean index obtained is $n=0.93\pm0.11$, in good agreement with the finding of 
\citet{won02} ($n=0.8\pm0.3$).  
Despite this overall agreement, there are slightly differences in the individual indexes $n$ for the five galaxies we have in common, mainly due to the fact that we exclude the innermost galaxy regions and that Wong \& Blitz choose a conversion  factor $X$ that does not vary as a function of the radius nor from one galaxy to another.\\
We checked that there is no evident correlation 
between the star formation rate and the atomic hydrogen, especially in the inner part of the galaxy disks 
where the H$\alpha$ emission is often strong but usually there is a low abundance of HI.
To study the role of the total gas in the Schmidt law
we compute the surface density profiles $\Sigma_{HI+H_2}$, combining the atomic and the molecular hydrogen profiles without any extrapolation.
In Figure \ref{schloch} (right panel) we plot the Schmidt law for the total gas surface density
on the same scale used for the molecular hydrogen alone (left panel).
Figure  \ref{schloch} shows that the addition of the atomic hydrogen
makes the correlation between the SFR and the gas better than considering the H$_2$ alone; the most striking feature is that the faintest points in the local Schmidt law that deviate the most when considering the molecular hydrogen alone, now follow a single power law.
Furthermore, we notice that the dispersion in the inner part of the disk is on average reduced, perhaps because by adding the atomic component we are taking into account the HI produced by the photodissociation in regions that have a high star formation activity.
Inspecting individual galaxies, we find that all but two galaxies with $r>0.8$ can be fitted using a single
power law with an rms $< 0.15$, i.e. the 60\% of the entire sample can be described with a single Schmidt law.
Using a least square fit, we derive a mean index $n=1.39\pm0.10$, in good agreement with the value obtained by Kennicutt using integrated quantities. Despite confirming the correlation between the SFR and the H$_2$
that was suggested by Wong \& Blitz, the important difference between our results and theirs is that 
the addition of HI improves this correlation, significantly reducing the number of galaxies for which the Schmidt law fails in the outer part of the disks.
Therefore we cannot agree with the conclusions of Wong \& Blitz that the correlation
for the total gas is entirely driven by the molecular component. A possible cause for this disagreement might be their use of CO imaging from the BIMA survey. Inspecting the H$_2$ radial profiles for the five galaxies we have in common, we detect molecular hydrogen emission that is up to two times more extended than in the BIMA observations.
The analysis of Wong \& Blitz seems more limited to the inner part of the galaxy disks, for which we agree 
that the correlation is almost entirely driven by the molecular component. Due to the superior sensitivity 
of the Nobeyama mapping to large-scale emission, we can push this analysis to the outer parts, where we find that the HI is the most important quantity that drives the correlation between the star formation rate and the total gas.
 
\begin{figure*}[t]
\centering
\begin{tabular}{c c} 
\includegraphics[width=8cm]{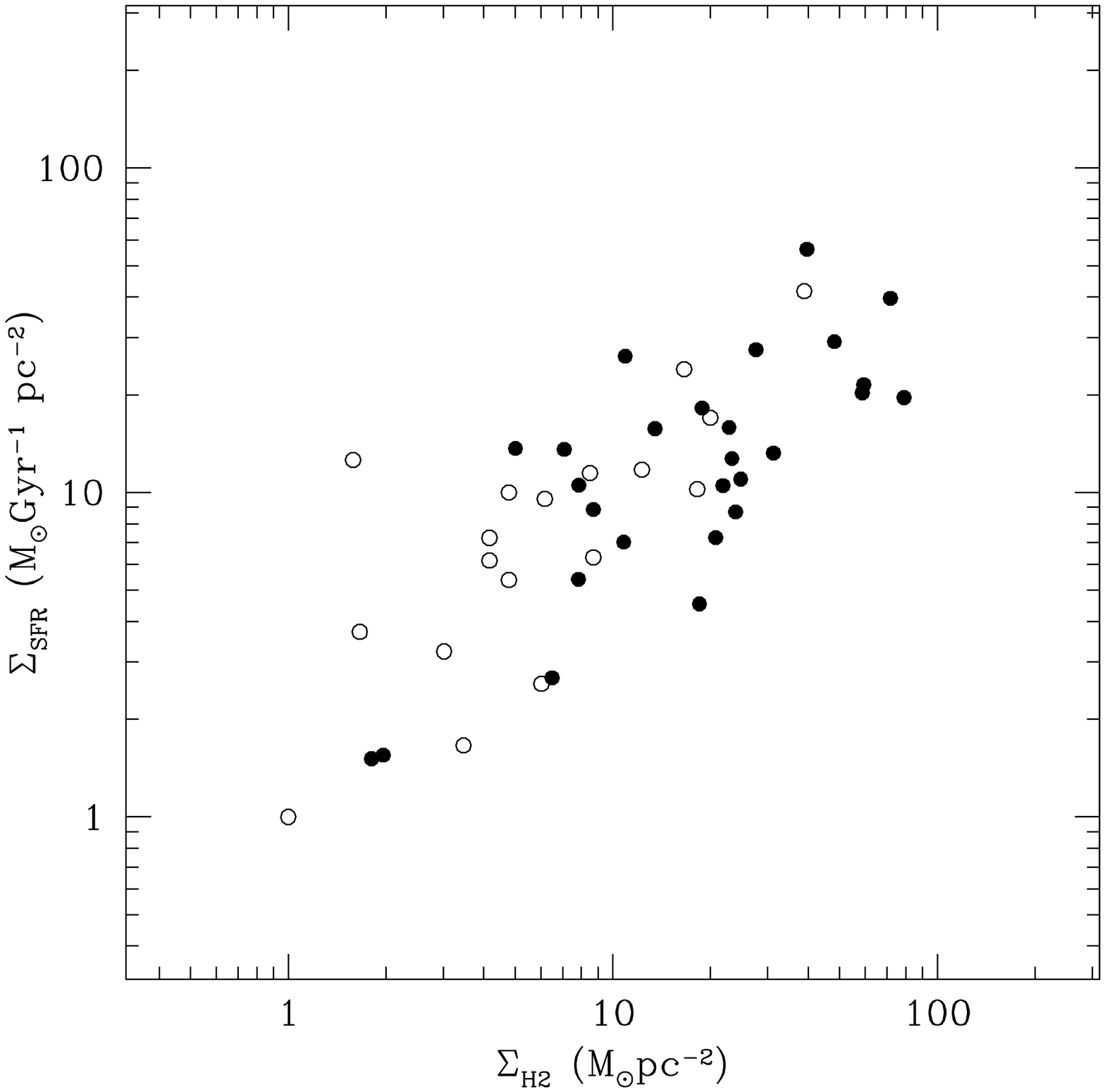}&\includegraphics[width=8cm]{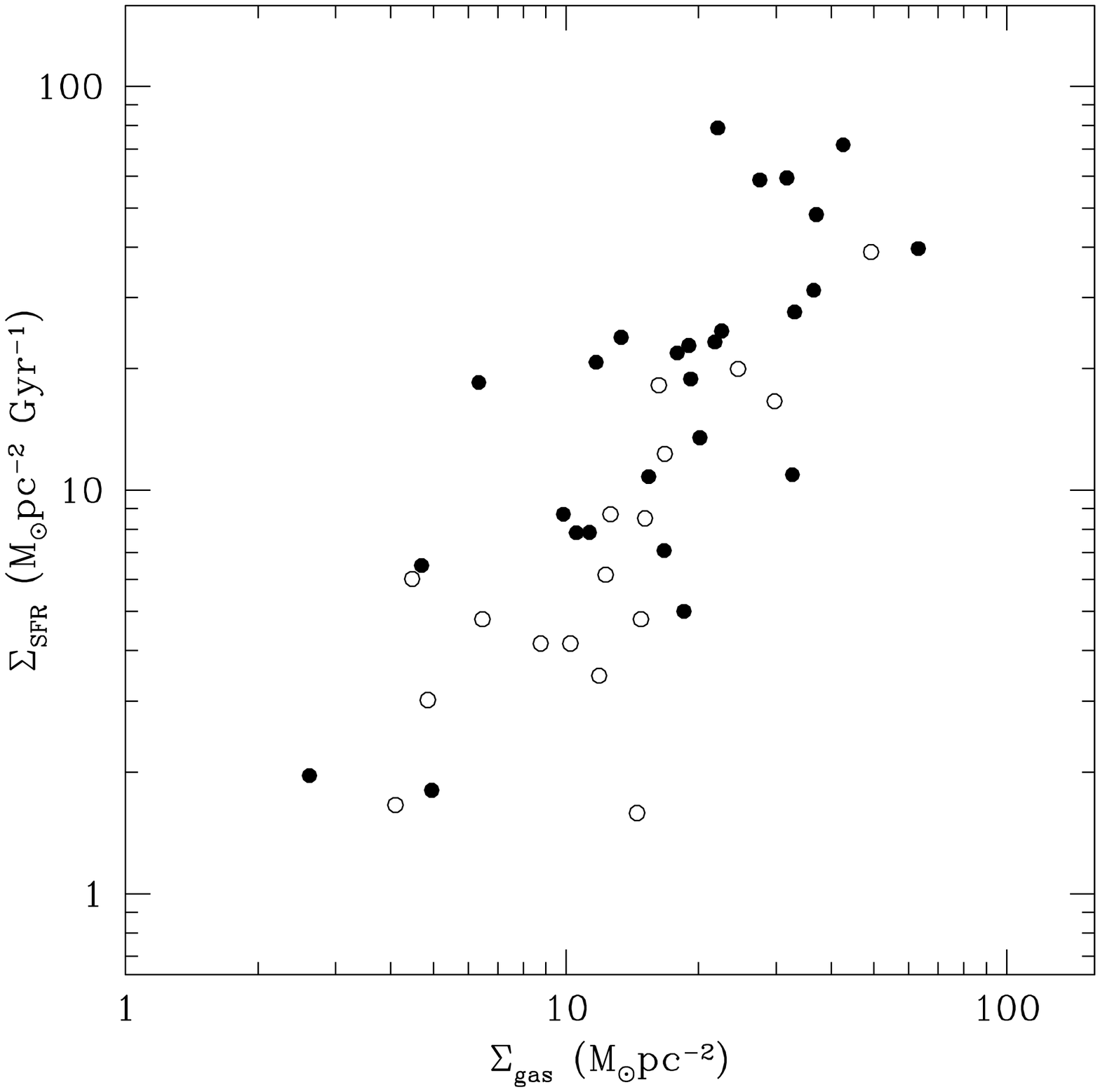}\\
\end{tabular}
\caption{The Schmidt law for integrated quantities, separately for the molecular (left panel) and the total hydrogen (right panel). Filled circles are for our measurements, while empty circles are data from \citet{ken98},
for the galaxies we have in common. Uncertainties are 0.2 dex for SFR and up to 1 dex for H$_2$ and for the total gas.}\label{globalsch}
\end{figure*}

\subsubsection{The global Schmidt law}
After having studied the local Schmidt law, we focus on its applicability while considering integrated quantities; therefore, we compute for each galaxy the total SFR and the gas surface density, by integrating
along the profiles.
In Figure \ref{globalsch} we show the correlation between the SFR and the H$_2$ (left panel)
and the correlation between the SFR and the HI+H$_2$ (right panel), for our measurements (filled circles) and for the galaxies we have in common, using data from \citet{ken98} (open circles). Uncertainties 
are not drawn, but as previously discussed the SFR has a typical error of 0.2 dex while the uncertainty
for the H$_2$ may be as large as 1 dex; the uncertainty for the total gas is entirely driven by that
of the molecular hydrogen.
We see that there is a fairly good agreement between our values and the ones from Kennicutt, confirming
the validity of our calibration procedures, in spite of the slightly different corrections.  
Considering the left panel, we find a good correlation ($r=0.76$) between the SFR density and the molecular hydrogen density; therefore, for our sample composed of massive galaxies, we can assume that
even on a global scale the Schmidt law is a suitable parametrization for the relationship between the 
SFR and the molecular gas.
Adding the HI (right panel), we can see the same good correlation ($r=0.77$) between the SFR density and the total hydrogen density; therefore, considering a sample of massive galaxies, we find that the Schmidt law for
the total gas does not differ significantly from the one for H$_2$ alone, thus the bulk of the star
formation that takes place in the inner part of the disks correlates with the molecular hydrogen
surface density, in agreement with what is found on local scales. This finding supports the hypothesis that 
the poor correlation with the H$_2$ found by Kennicutt is driven by the inclusion of low luminosity galaxies
for which it is difficult to obtain reliable CO observations and for which a constant $X$ conversion
factor does not hold.

\subsection{The environmental effects}
 
In the previous section we have shown that the atomic hydrogen seems to play some role in the star formation process, especially at the edge of galaxy disks; moreover, the atomic hydrogen, built up during the 
cosmological nucleosynthesis, is the primeval constituent of the molecular hydrogen and therefore
provides the essential fuel for the star formation, even if indirectly.
Due to this fact, we now focus on the possible perturbations to the star formation activity 
that might be induced by the removal of the neutral hydrogen in  the environment of a rich cluster of galaxies.
The most suitable parameter to quantify the HI removal in a galaxy is
the HI deficiency, computed according to \cite{hay84} as the logarithmic difference between $M_{HI}$ of a reference sample of isolated galaxies and $M_{HI}$ actually observed in individual objects: 
$def_{HI}= \log M_{HI~ref} - \log M_{HI~obs}$. 
$\log M_{HI~ref}$ has been found to be linearly correlated to the galaxy linear diameter $d$ as 
$\log M_{HI~ref}=a+b \log d$, where $d$ (kpc) is determined at the $25^{th}$ B--band isophote,    
and $a$ and $b$ are weak functions of the Hubble type. Here we adopt the updated coefficients 
from \citet{sol96}. In the following analysis we choose to consider a galaxy HI deficient if the ratio between the measured and the expected HI mass is $>2.5$, i.e. if $def_{HI}\geq0.4$. Therefore, our sample is composed of 54\% non deficient galaxies that we consider unperturbed and 46\% deficient galaxies that suffer from significant HI removal.

\subsubsection{Global abundances}

\begin{figure*}[t]
\centering
\begin{tabular}{c c}
\includegraphics[width=8cm]{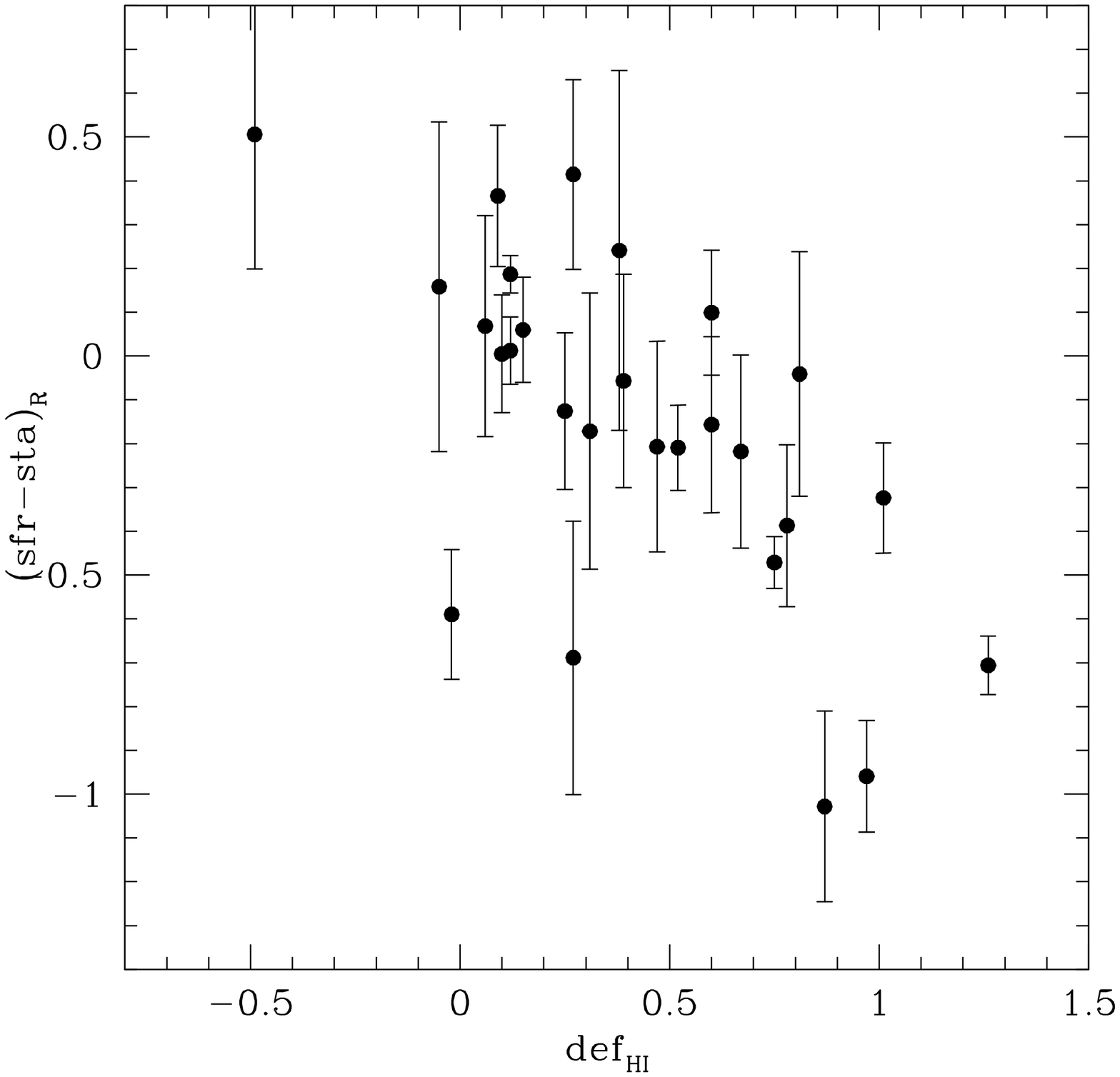}&\includegraphics[width=8cm]{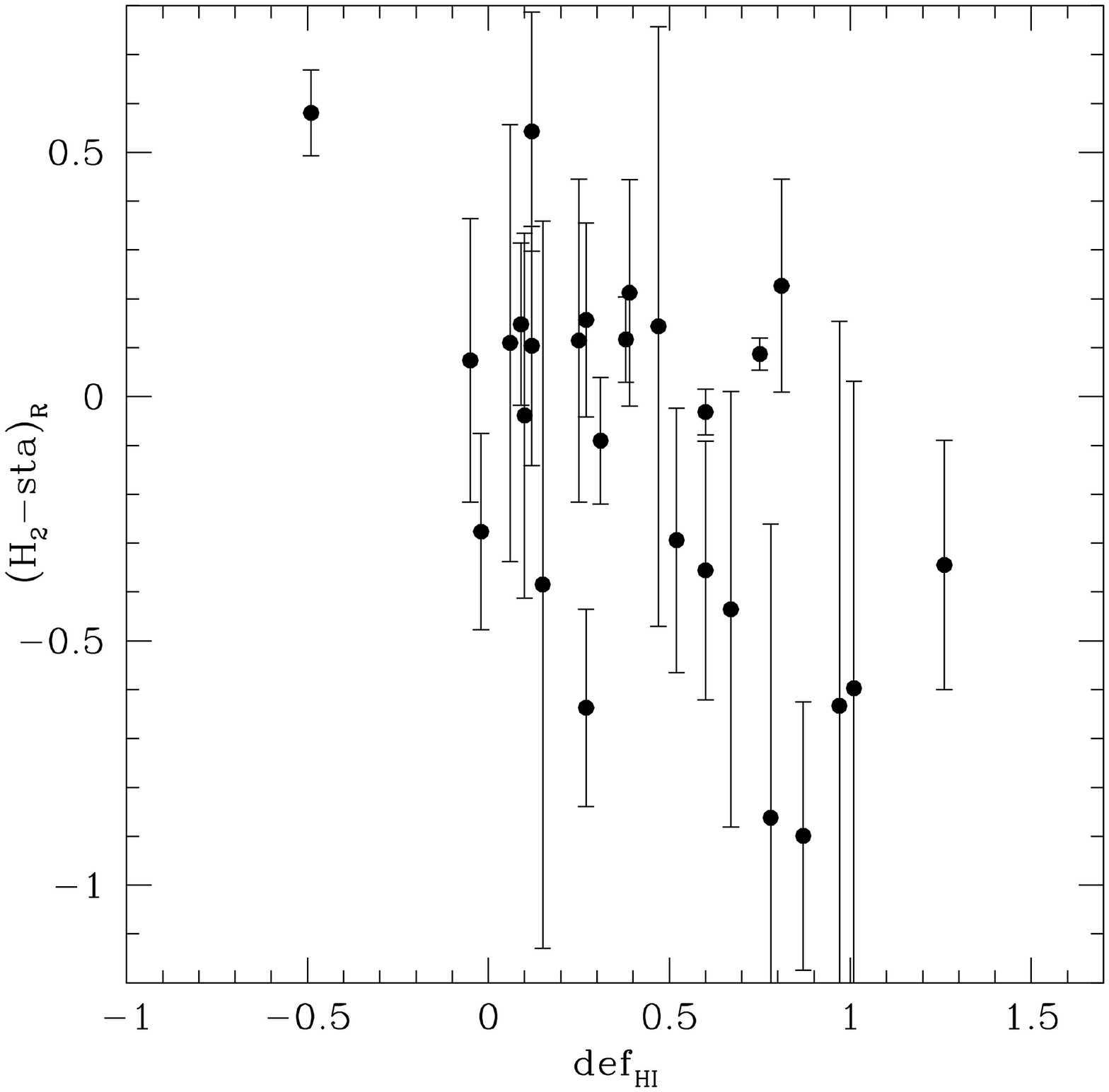}\\
\end{tabular}
\caption{The relationship between $def_{HI}$ and color residuals $(H_2-sta)_R$ (right panel) and 
$(sfr-sta)_R$ (left panel). Correlation coefficients are respectively $r=0.60$ and $r=0.65$. Uncertainties are  discussed in the text.}\label{global}
\end{figure*}

Firstly we focus on the relationship between the global SFR, H$_2$ mass and stellar mass
as a function of the deficiency.  
Therefore, for each galaxy we compute a mean color along the radial profile between three pairs of quantities:
$sfr-sta$, $sfr-H_2$ and $H_2-sta$, averaged over the entire disk.
Since the color profiles typically exhibit a radial gradient that in some cases can be significant, considering
a mean value for each galaxy may sometimes be a too crude assumption.
We quantify this effect calculating the standard deviations $\sigma$ and we obtain:\\
$0.07\leq\sigma\leq0.40$ with a mean of $\bar\sigma=0.20$ dex for $sfr-sta$,\\  
$0.05\leq\sigma\leq0.40$ with a mean of $\bar\sigma=0.25$ dex for $sfr-H_2$ and \\  
$0.03\leq\sigma\leq0.60$ with a mean of $\bar\sigma=0.31$ dex for $H_2-sta$; \\ these deviations,
plotted in Fig.\ref{global}, provide an estimate of the uncertainty in addition to the errors in the profile calibrations.
To clarify the physical meaning of these colors we notice that $sfr-sta$ is similar to the H$\alpha$ equivalent
width. 
In fact the $sfr-sta$ color and the H$\alpha$ equivalent width (E.W.)
(retrieved from GOLDMine or collected from \citet{koo04,rom90,ken83d}) are found to be tightly correlated ($r=0.84$). Similarly the $H_2-sta$ color quantifies the H$_2$ emission, normalized to the stellar continuum; finally $sfr-H_2$ can be considered a tracer of the efficiency at which the molecular
gas is converted into new stars.
It is known that galaxies obey empirical scaling laws according to 
their dynamical mass and therefore to their H--band luminosity, owing to the tight 
correlation between these two quantities \citep{gav96b}. 
As the luminosity increases, the HI mass to  
luminosity ratio M$_{HI}$/L$_{H}$   \citep{gav08}, the H$_2$ mass to  
luminosity ratio M$_{H_2}$/L$_{H}$  \citep{bos97} and the H$\alpha$ equivalent width \citep{gav96b}
decrease; in other words, the smallest galaxies are, relative to their mass,
the richest in gas and the most active in star formation.
Before studying if the above three sets of colors display a residual dependence on $def_{HI}$ independently of their mass, we need to correct for the primary luminosity dependence.
Unperturbed galaxies in our sample ($def_{HI}<0.4$) obey the following relationships, derived using a 
least square fit:
\begin{equation}
\rm H_2-sta=-0.25\times \log L_H + 1.55 ~(rms=0.30 dex)
\end{equation}  
\begin{equation}
\rm sfr-sta=-0.26\times \log L_H + 1.95 ~(rms=0.32 dex).
\end{equation}  
In Figure \ref{global} we plot, as a function of the HI deficiency, the residuals $(H_2-sta)_R$ (left panel) and 
$(sfr-sta)_R$ (right panel) computed using the previous equations.
Inspecting the left panel, we can see a significant correlation ($r=0.65$) between the $(sfr-sta)_R$ and the deficiency;
since environmental processes do not significantly affect the stellar distribution in massive galaxies, the meaning of this correlation is that HI removal causes an increasing reduction in the SFR with increasing deficiency.
In the right panel a weaker but still significant correlation ($r=0.60$) exists  between  
$(H_2-sta)_R$ and the $def_{HI}$, showing that molecular hydrogen, even if well
bounded in the galaxy potential well, can be perturbed by the HI removal, at least indirectly.
Conversely we find (not shown) that $sfr-H_2$ and $def_{HI}$ are uncorrelated quantities ($r=0.20$), suggesting that the efficiency of molecular
gas to star conversion does not depend significantly on the amount of HI removal.

\subsubsection{Radial behavior}
\begin{figure*}[t]
\centering
\begin{tabular}{c c c}
\includegraphics[width=6cm]{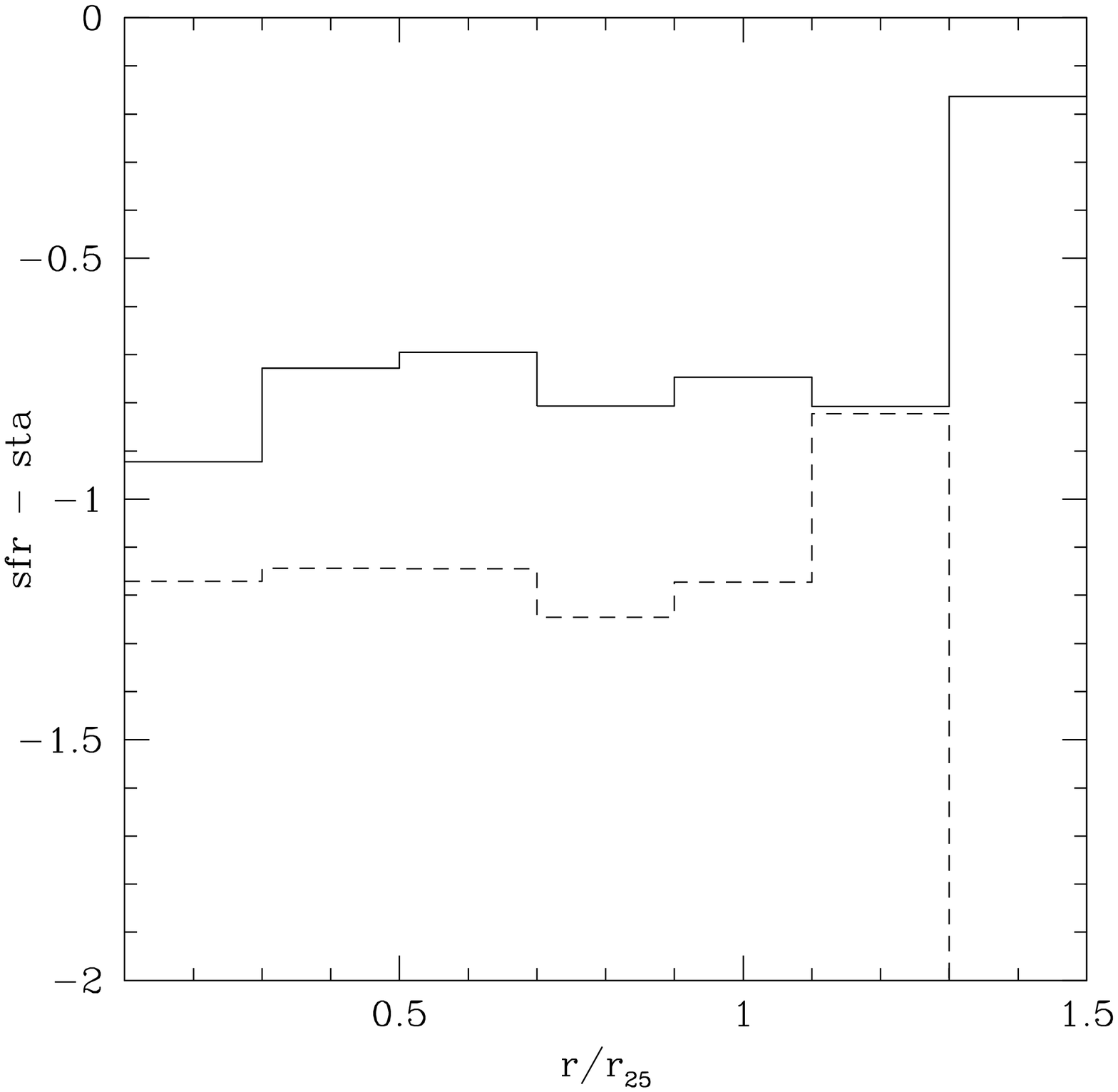}&\includegraphics[width=6cm]{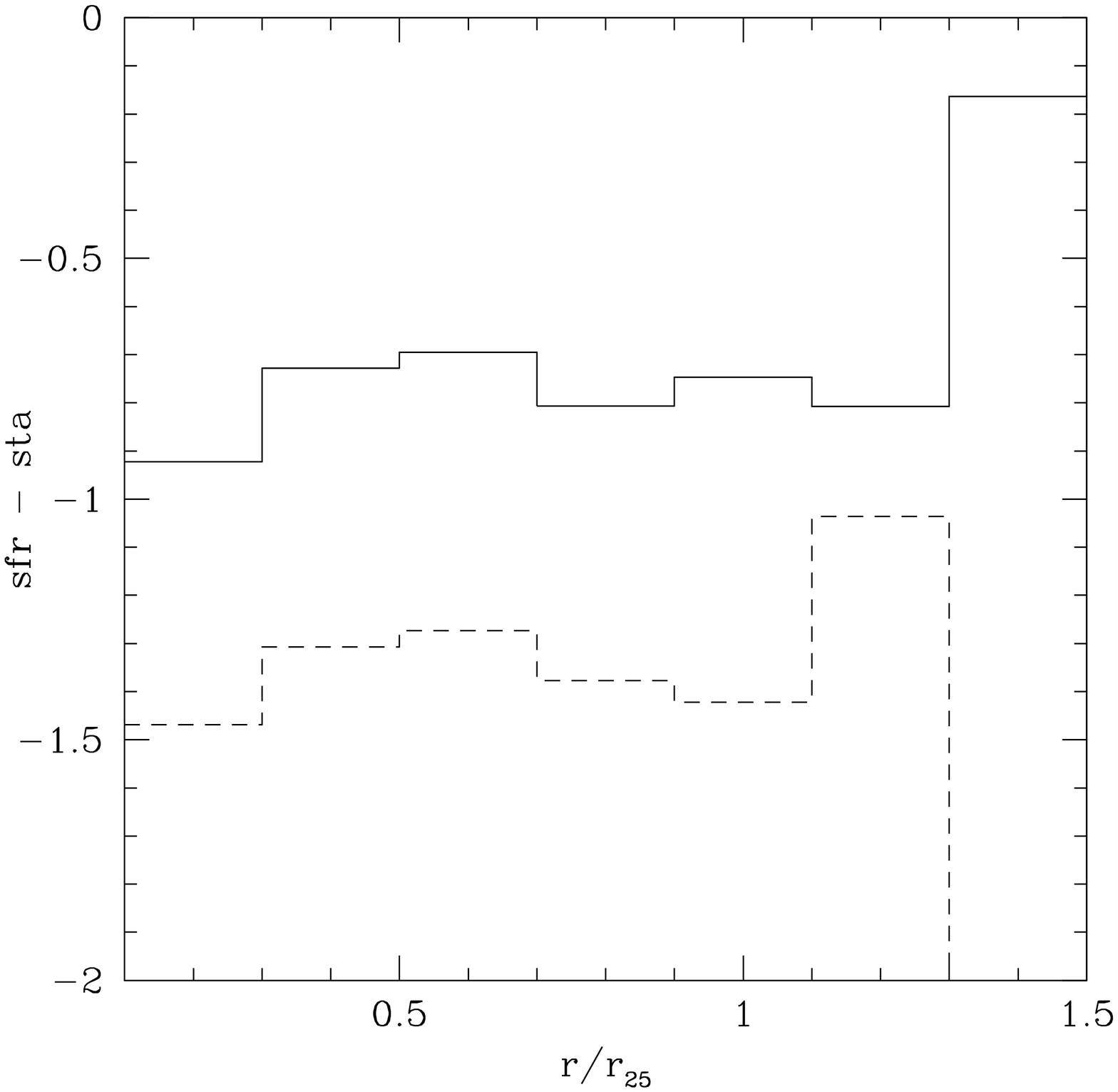}&\includegraphics[width=6cm]{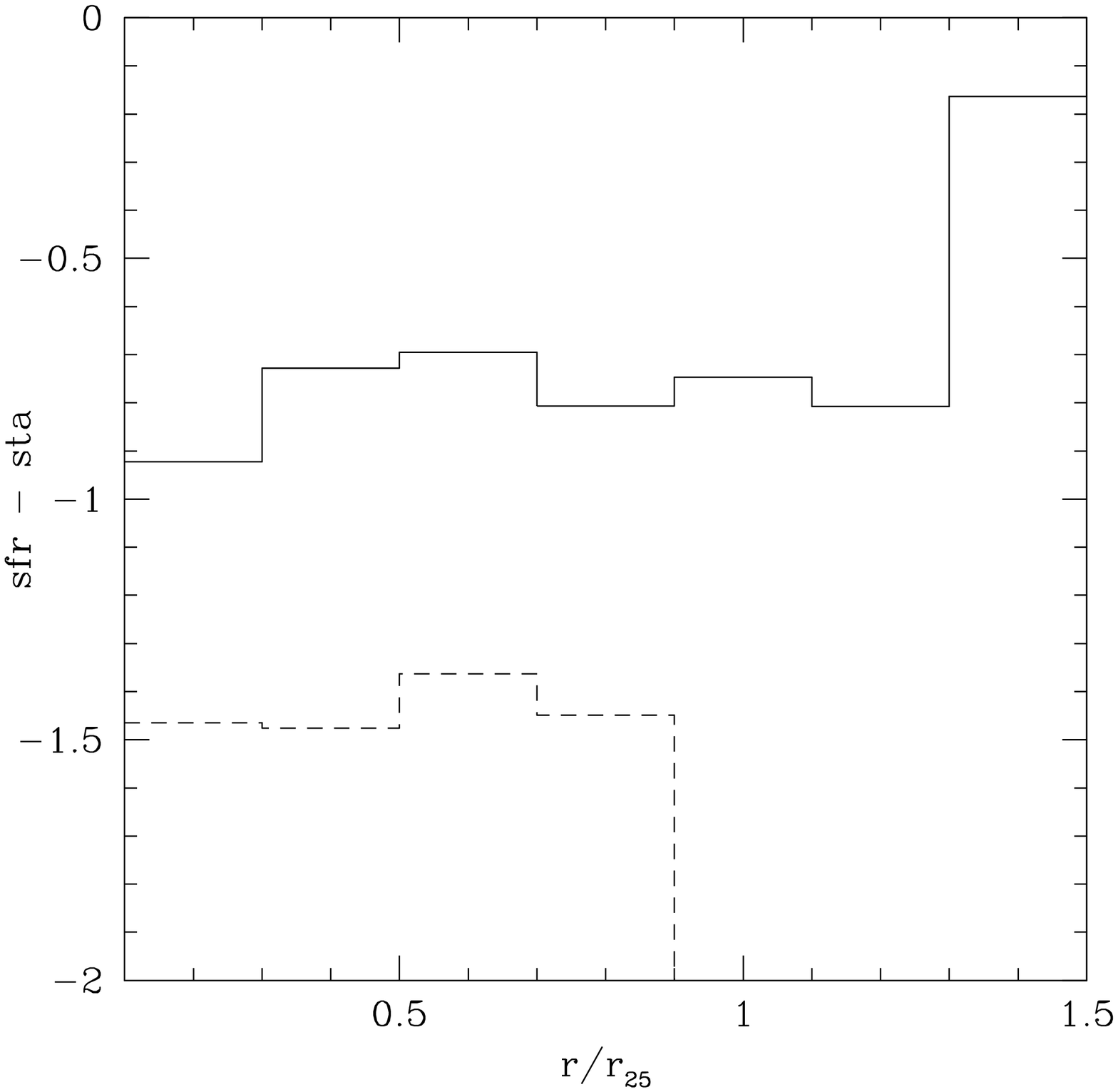}\\
\end{tabular}
\caption{Comparison between the mean $sfr-sta$ profile for deficient (dashed lines) and non deficient 
(continuous lines) galaxies; from left to right, deficiency threshold is set to be $def_{HI}\geq0.4$, 
$def_{HI}\geq0.65$ and $def_{HI}\geq0.9$. Non deficient galaxies are always considered for $def_{HI}<0.4$.}\label{smootsfr}
\end{figure*}

After having shown that both the SFR and the molecular abundance
are reduced in deficient galaxies on a global scale, a new question arises: is this  reduction
mainly due to a global quenching or it does take place primarily in the external disk, producing a truncation in the star forming disk?
This issue has been addressed, among others, in a series of studies by Koopmann et al. They find that
Virgo galaxies have their H$\alpha$ disk truncated, while the case of starvation
that produces a global reduction in star formation across the entire disk is rare \citep{koo04}.
To assess this result in our sample and to verify if similar conclusions hold also for the molecular
gas, we study the radial behavior of the color profiles.
We normalize the radius of each galaxy using r$_{25}$ from Table \ref{sample}
and we compute the mean colors in bins of r/r$_{25}$, averaging over all galaxies in each bin.
In Figure \ref{smootsfr} we show a comparison between the mean $sfr-sta$ profile for deficient 
(dashed lines) and non deficient (continuous lines) galaxies; from left to right, the deficiency
threshold increases, i.e. in the left panel deficient galaxies are considered with $def_{HI}\geq0.4$,
in the middle those with  $def_{HI}\geq0.65$ and in the right panel the ones with $def_{HI}\geq0.9$. Conversely, non deficient galaxies have $def_{HI}<0.4$ in all panels.
The first panel shows a gap of 0.4/0.5 dex between the $sfr-sta$ color for non
deficient and deficient galaxies, suggesting that the SFR in deficient galaxies is on average 
reduced. 
\begin{figure*}[t]
\centering
\begin{tabular}{c c c}
\includegraphics[width=6cm]{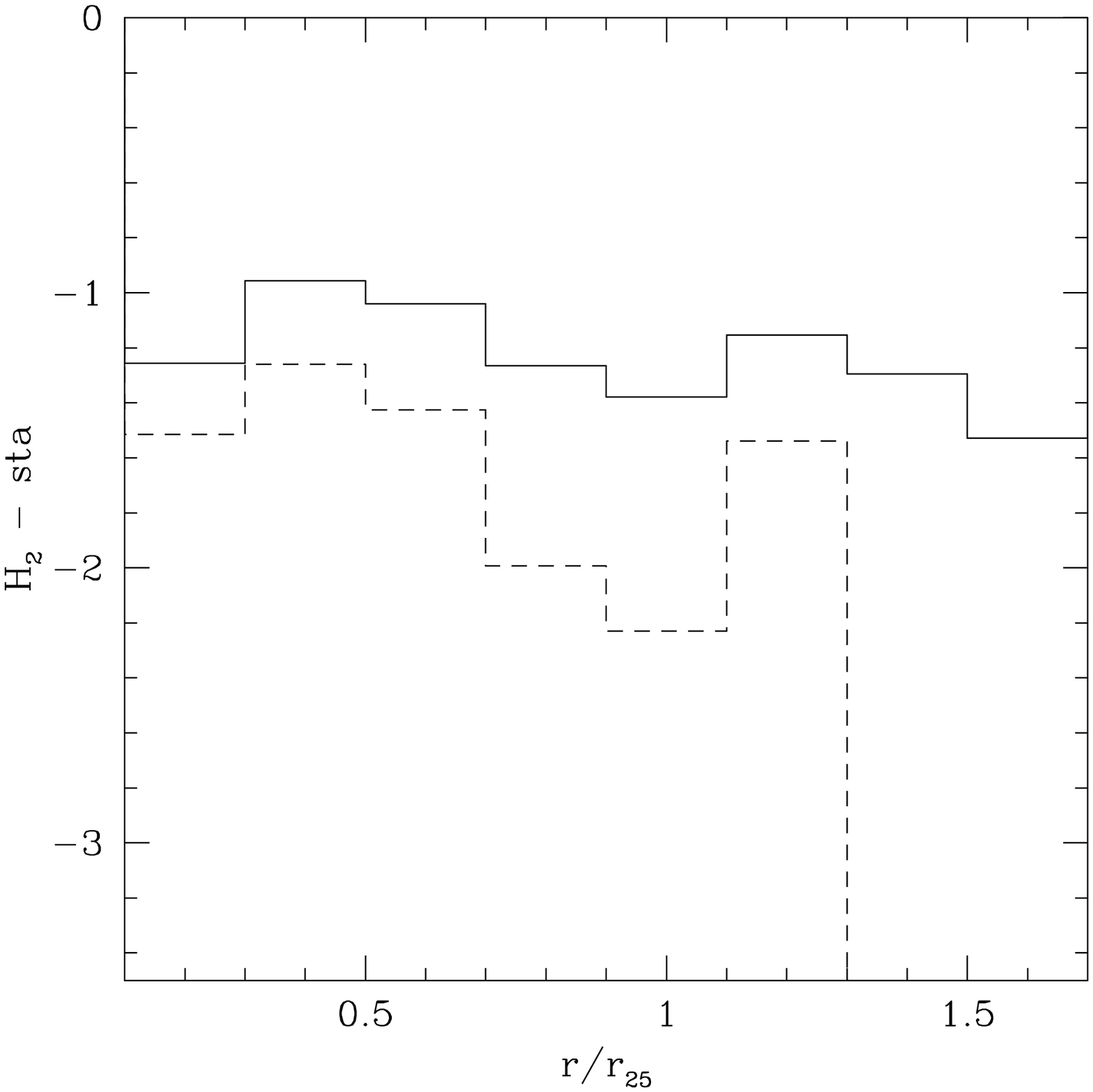}&\includegraphics[width=6cm]{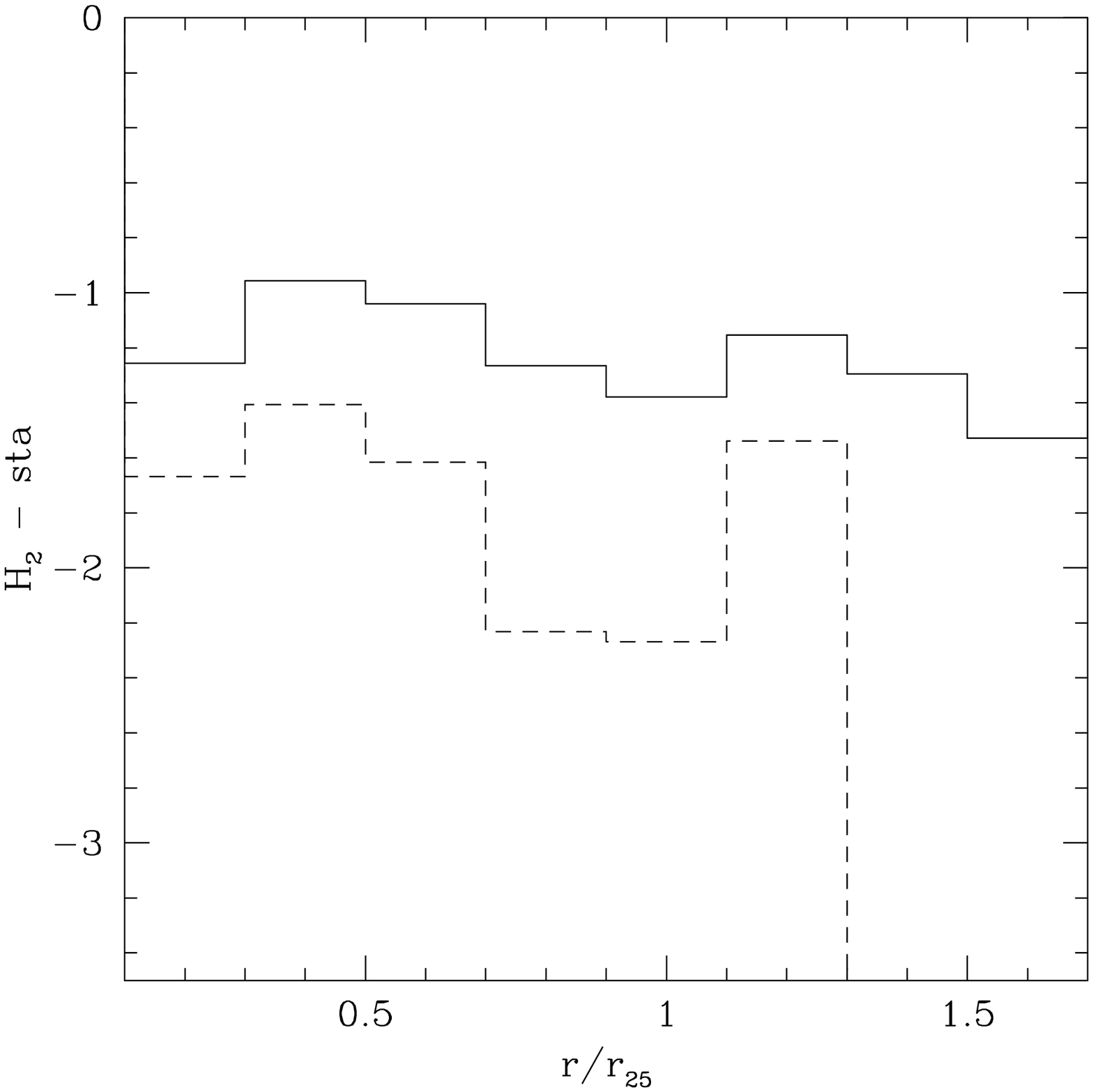}&\includegraphics[width=6cm]{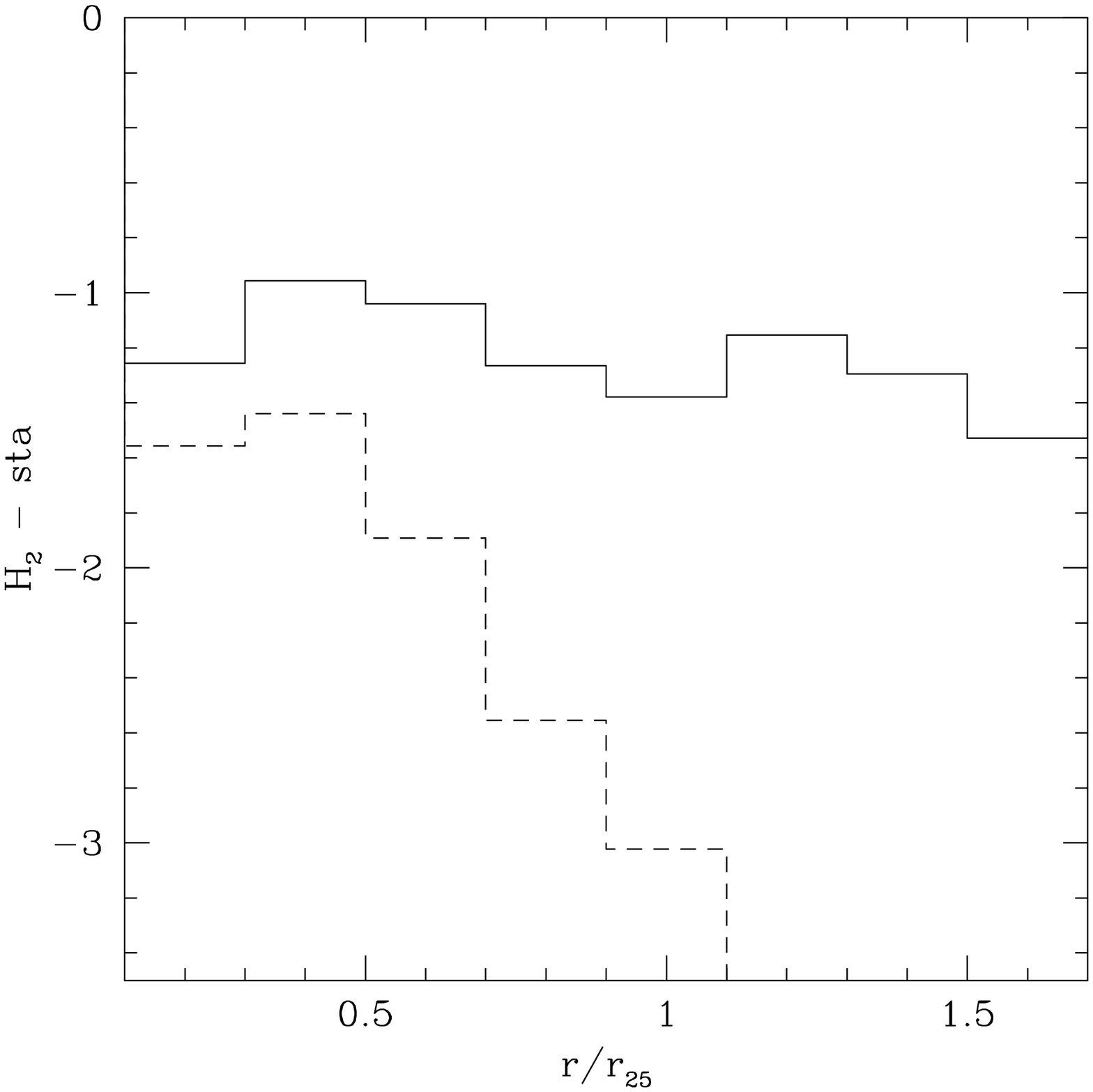}\\
\end{tabular}
\caption{Comparison between the mean $H_2-sta$ profile for deficient (dashed lines) and non deficient 
(continuous lines) galaxies; from left to right, deficiency threshold is set to be $def_{HI}\geq0.4$, 
$def_{HI}\geq0.65$ and $def_{HI}\geq0.9$. Non deficient galaxies are always considered for $def_{HI}<0.4$.}\label{smootco}
\end{figure*}
Increasing the deficiency threshold (middle panel)
the gap increases up to 0.6 dex, while for the highest deficient galaxies
$def_{HI}\geq0.9$  (right panel) the gap increases up to 0.8 dex. 
Moreover, there are no galaxies with H$\alpha$ emission beyond r/r$_{25}=0.8$; therefore, for extremely deficient galaxies, besides the global quenching of the star formation rate,  there is an evident truncation effect.
This result suggest that the correlation found between the global $sfr-sta$ and the $def_{HI}$ is mainly
driven by quenching, while truncation is important only for highly deficient galaxies.
For $H_2-sta$, an effect similar to the one described for the $sfr-sta$ is present
in Figure \ref{smootco}.
In fact, in the left panel the gap between non deficient and $def_{HI}\geq0.4$ galaxies is 0.4/0.6 dex, it
increases to 0.8 dex when $def_{HI}\geq0.65$ (middle panel); for $def_{HI}\geq0.9$ (right panel) the gap 
increases significantly with increasing radii and there is no emission beyond
r/r$_{25}=1$. Once again, this finding is consistent with the fact that the correlation found between 
the global $H_2-sta$ and the $def_{HI}$ is mainly driven by quenching, while truncation is important for highly deficient galaxies only.
Conversely we find no significant difference between non deficient and deficient galaxies as far
as the radial variation of the $sfr-H_2$  efficiency is concerned, being on average within
$\pm$ 0.2 dex.
\begin{figure}[t]
\centering
\includegraphics[width=8cm]{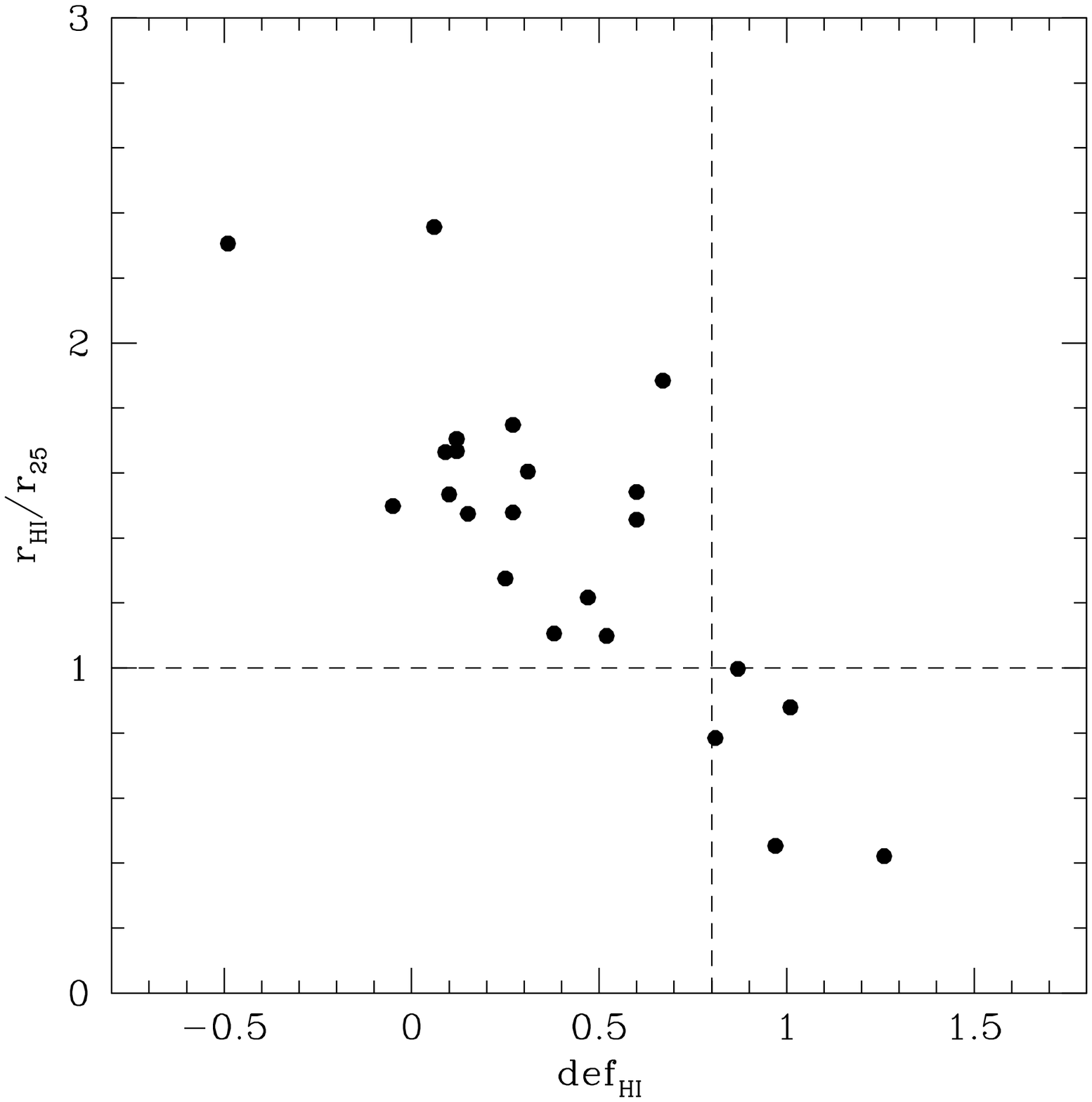} 
\caption{The relationship between $def_{HI}$ and the HI isophotal radius at $5 \times 10^{19}$ cm$^{-2}$, normalized to the r$_{25}$. Correlation coefficient is $r=0.81$. The  dashed lines emphasize that
r$_{HI}>$r$_{25}$ for $def_{HI}>0.9$.}\label{hirad}
\end{figure}
\begin{figure}[t]
\centering
\begin{tabular}{c}
\includegraphics[width=8cm]{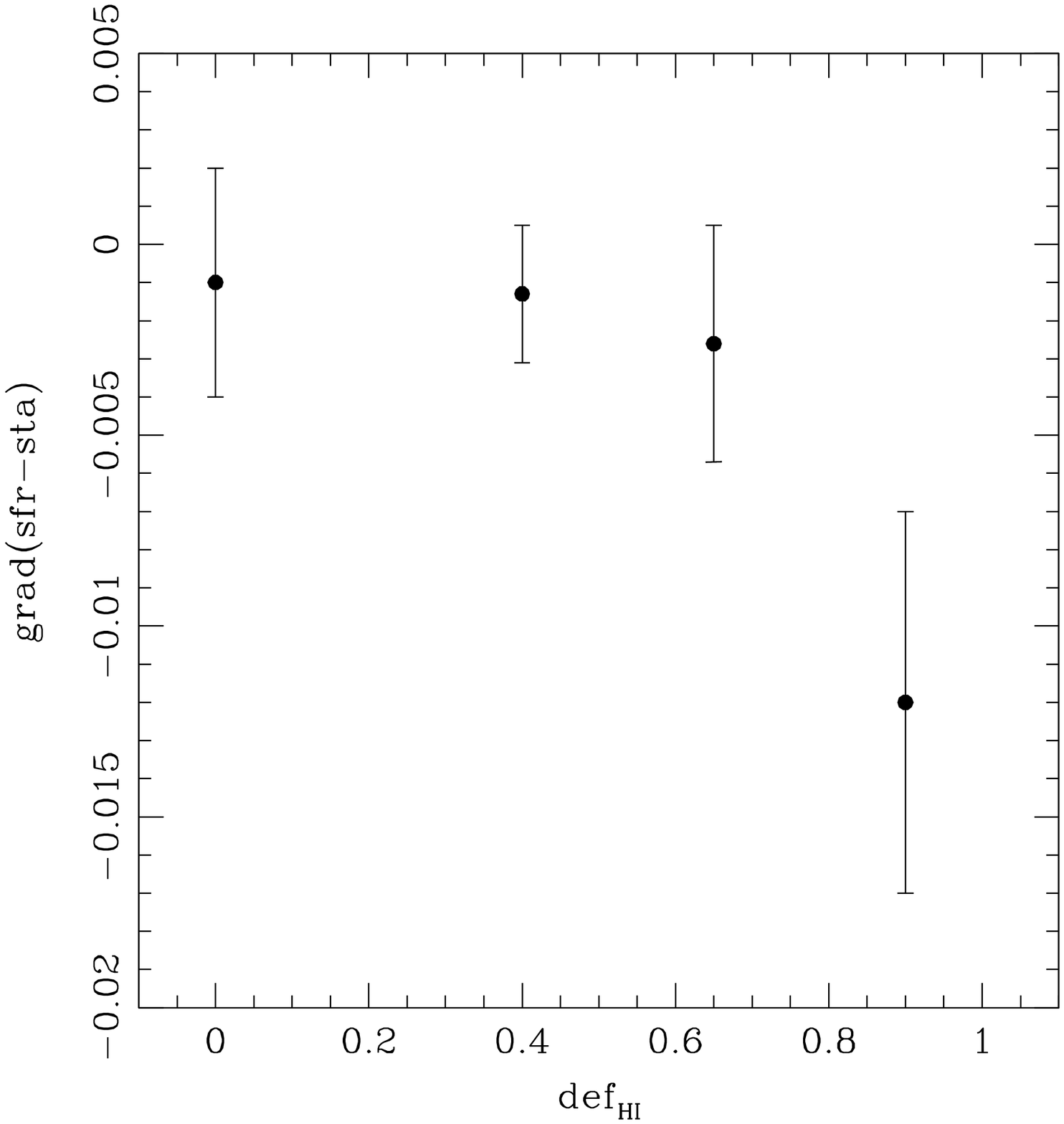}\\
\includegraphics[width=8cm]{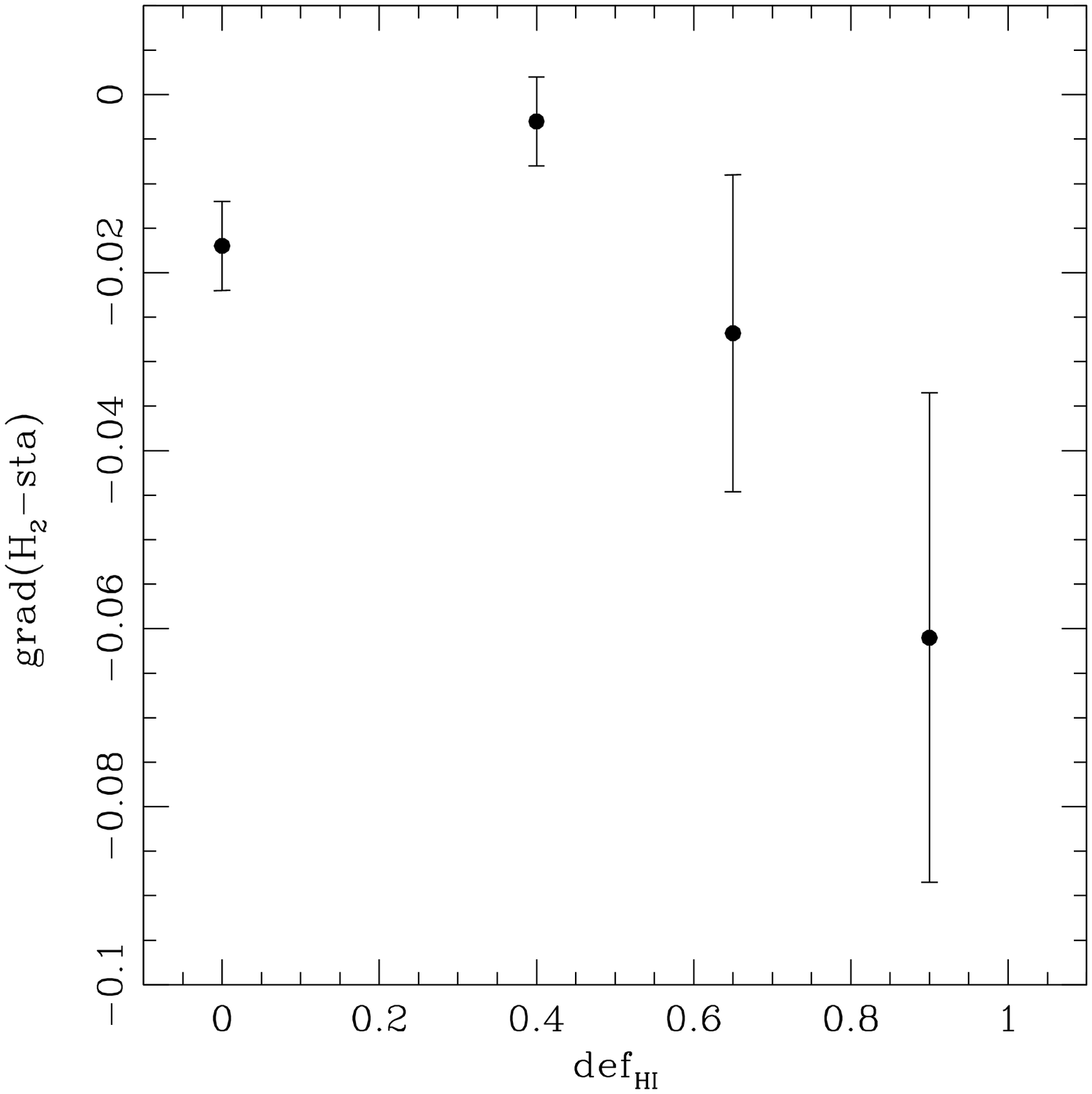}\\
\end{tabular}
\caption{Mean radial gradients for the outer part of the $sfr-sta$ (top panel) and $H_2-sta$ (bottom panel) 
colour profiles, in four bins of deficiency (0-0.4, 0.4-0.65, 0.65-0.90, $>$0.90). Error bars are the uncertainties on the mean values in each bin.}\label{grad}
\end{figure} 
The existence of a significant correlation  between the extent of the HI disk and the $def_{HI}$ parameter
is a well known result \citep[e.g.][]{cay94} implying that environmental effects, and in particular 
ram pressure stripping, perturb the ISM distribution outside-in, as also found in the simulations \citep{aba99}. 
In this work we consider the HI isophotal radius, taken arbitrarily at 
$5 \times 10^{19}$ cm$^{-2}$, as a suitable measure of the HI disk. This threshold well matches the sensitivity of most of our HI data, except for NGC 3521, 4192, 5055, 5236 and 5457 where the HI column density stays
above the threshold at all radii, but their exponential decay can be confidently extrapolated outside; 
only for NGC 4402 and NGC 4579, whose HI data are limited to the very inner regions, would any extrapolation 
be meaningless and we exclude them.
In Figure \ref{hirad} we show the tight correlation ($r=0.81$) between the 
isophotal HI radii normalized to r$_{25}$ and $def_{HI}$, allowing us to quantify that 
at very large deficiencies ($def_{HI}\geq0.9$) it occurs that 
the extent of the HI disk becomes smaller than the optical radius.
Is there something fundamental in the transition between galaxies that have their HI disk
smaller or bigger than the optical radius? Does this transition also govern 
the difference in the SFR and H$_2$ radial profiles between extremely deficient and normal galaxies?
To test these hypotheses, we further investigate if the outer parts of the SFR and H$_2$ radial profiles 
exhibit a different behavior as a function of the deficiency. 
We derive for each galaxy the radial gradient by fitting the color profiles ($sfr-sta$ and $H_2-sta$) in the outermost regions (r/r$_{25}\geq0.50$). 
The fit is performed with a maximum likelihood method, assuming a constant uncertainty along the profiles and using
an exponential, a reasonable model for the $H_2-sta$ since both the molecular and the stellar component typically behave as exponential disks.
The issue is more complex for the $sfr-sta$ since the H$\alpha$ emission
tends to be rather erratic on small scales; however, after a 15 arcsec smoothing
even the H$\alpha$ profiles can be fitted with an exponential function.
The results of this analysis are summarized in Figure \ref{grad}, where we plot against $def_{HI}$ the mean radial gradients 
for the outer part of the $sfr-sta$ (top panel) and $H_2-sta$ (bottom panel) color profiles;
the mean values are derived averaging in four bins of deficiency (0-0.4, 0.4-0.65, 0.65-0.90, $>$0.90)
and the error bars refer to the uncertainties on these averages.
The first panel of Figure \ref{grad} shows that the mean gradients of the 
$sfr-sta$ color profiles are similar for non deficient and moderately deficient galaxies 
(up to $def_{HI}=0.90$), while when considering highly deficient galaxies the discrepancy becomes significant at $1\sigma$.
Finding steeper $sfr-sta$ profiles for highly deficient galaxies implies that when the
$def_{HI}>0.9$ and thus the HI disk is reduced inside the optical radius, the SFR sharply
decreases in the outer part of the disk, consistently with the truncation effect.   
The second panel shows that the discrepancy between the mean gradients of the 
$H_2-sta$ color profiles in non deficient or moderately deficient galaxies and extremely deficient galaxies
is less sharp. In fact, due to the great dispersion in the 0.65-0.90 and $>$0.90 bins, the
difference between highly deficient galaxies and moderate deficient galaxies is less marked; however, it is 
evident that the steepest $H_2-sta$ are found in galaxies with $def_{HI}>0.9$, once again supporting 
the idea that also the H$_2$ profiles sharply decrease in the outer part of the highly deficient disks, 
suggesting that the truncation may also affect the H$_2$ distribution.

\section{Discussion}\label{results}
\subsection{The Schmidt law}
The study of the local Schmidt law addresses one of the main
open questions about the star formation activity in spiral galaxies: is the process of star formation related
to the atomic or the molecular component? Clearly, the formation of stars takes place
inside the GMCs, in cold and dense molecular regions on a scale of few parsecs; however,
while studying the molecular distribution in our sample at a resolution of $15''$, our observations are sensitive
to the diffuse component on typical scales of 1 kpc; this molecular medium can 
sometimes be not self gravitating and therefore it may not take part to the star formation
activity \citep{elm02}.
From our analysis of the local Schmidt law we find that there is a good correlation
between the SFR and the H$_2$, that eventually becomes poorer in the outer part of the disks (see Fig. \ref{schloch}, left).
Since the bulk of the star formation activity takes place
in the inner disks, it can be concluded that the Schmidt law for the molecular component 
is a good model of the SFR in galaxies.
Therefore it seems plausible that the molecular medium on the kpc scale 
traces the self gravitating molecular clouds which form at much smaller scales and sustain
the star formation.
However we found that the Schmidt law computed for the  total gas, i.e. including the atomic hydrogen,
is significantly improved, particularly in the outer disks (see Fig. \ref{schloch}, right).
This result supports the theoretical argument by \citet{elm02} that some clouds can be in a molecular state without being strongly self--gravitating and, on the contrary, there might be instabilities in the HI component
capable of triggering the star formation via direct collapse of HI; therefore it is reasonable to accept 
that the star formation rate scales with the total column density of gas in all forms, including a possible contribution from photodissociation of H$_2$ molecules.
We thus disagree with the conclusion of \citet{won02} who state 
that the correlation in the Schmidt law is driven entirely by the molecular component.
The fact that we find a mean index $n=1.4$ for the local Schmidt law, similar to the one originally proposed by Kennicutt 
using integrated quantities, seems to imply that this relation is scale invariant, i.e. it is able to describe the process of star formation on scales ranging from 1 kpc to the entire galaxy disk.

\subsection{Environmental effects}
The HI is the primary fuel for the star formation in galaxies, both directly and indirectly.
We have shown that at the disk periphery the HI might directly contribute to the star formation
by collapsing into molecular clouds and indirectly by providing the constituent of the diffuse H$_2$ 
medium, composed of molecules which form on dust grains inside the stellar potential wells \citep{bli04}.
Therefore, it is not surprising that the HI removal caused by the environmental processes may affect
the star formation activity in cluster galaxies;  we have shown that HI deficient
galaxies have their SFR and H$_2$ abundances reduced with respect to normal galaxies.
Given the fact that we find no significant dependence of the SF efficiency on the
environmental conditions, the observed trend of decreasing SFR and H$_2$ density in deficient galaxies
is consistent with models of galaxy $starvation$, as proposed by \citet{lar80} 
and elaborated by \citet{bek02}, 
according to which the effect of reducing the HI halo halts the 
replenishment of H$_2$, leading to a quenching of the star formation rate. 
This model implies a continuous HI inflow from outside, as observed in some nearby
galaxies, although at a rate that seems insufficient to sustain the current rate of star formation \citep{san08}.
When galaxies become extremely HI deficient, i.e. when the hostile environment has produced 
the ablation of their neutral gas up to a factor of ten and the HI disk reduces its size within
the optical radius, superimposed on starvation, significant truncation of the star formation appears
to exist, i.e. the SF disk is sharply eroded outside-in. 
The disagreement with \citet{koo04}, who maintain that the reduction in the global star 
formation in the Virgo cluster is caused primarily by the spatial truncation of the star--forming disks, and not due to starvation, is only suggested. Their 
sample is dominated by extremely deficient spirals (among their deficient galaxies, 
65\% have $def_{HI}\geq0.9$ and  85\% have $def_{HI}\geq0.8$). 
Beside the agreement with previous studies on the SFR in clusters, our analysis points out for the first
time that the H$_2$ content might be affected by environmental processes, even if indirectly as a consequence of  HI depletion.
     
\section{Conclusions}\label{concl}

With the aim of exploring the relations between the process of star formation and the physical
properties of the ISM in various galaxy environments, 
we collected state-of-the-art imaging material and maps for 28 massive spiral galaxies belonging
to the Virgo cluster and to the local field. 
The observational material includes: images of the stellar continuum (taken in the red/NIR bands);
H$\alpha$ images of the young ($<4\times 10^6$ yrs) stars; 
radio maps at 2mm from the recent {\it Nobeyama CO atlas of nearby spiral galaxies} that combines
good sensitivity to large-scale CO emission with a 15 arcsec spatial resolution, and
sensitive radio maps at 21 cm from the ongoing, yet unpublished, VIVA and THINGS surveys carried out at the VLA.
Physical parameters have been derived homogeneously and carefully for the individual galaxies
applying corrections based on measured quantities, rather than average values. 
We hope that the effort we put in the analysis of the data reduces the statistical limitations that
arise from the paucity of imaging material currently available.  \\
The present analysis indicates  
that the bulk of the star formation in spiral galaxies is
supported by a diffuse molecular medium which forms through the conversion of the 
atomic hydrogen, due to the pressure exerted by the stellar potential. 
However, at the edge of the star forming disks, the HI plays a more important role  
than previously believed in directly sustaining the star formation activity.
When environmental processes cause significant removal of the outer part of 
the HI disks, galaxies suffer from starvation; the replenishment of the atomic gas 
inside the optical disks is reduced, leading to a depletion of the molecular component that is
consumed during the star formation activity, thus causing the quenching of the star formation activity itself.
When the HI removal is so severe that the HI disk shrinks inside the optical radius, 
there is also a truncation of both the molecular and the star forming disk.

\bibliographystyle{aa} 
\bibliography{biblio.bib} 

\appendix
\begin{acknowledgements}
We thank A. Leroy, E. Brinks, A. Chung, J. van Gorkom and J. Kenney for their kind permission to use HI maps
from THINGS and VIVA, prior to publication.
We thank A. Boselli, L. Cortese, C. Bonfanti, P. Franzetti and B. Devecchi for useful hints
and discussions.
This research has made use of the GOLDMine Database.
This research has made use of the NASA/IPAC Extragalactic Database (NED) which is operated by the Jet Propulsion Laboratory, California Institute of Technology, under contract with the National Aeronautics and Space Administration.
We acknowledge the usage of the HyperLeda database (http://leda.univ-lyon1.fr).
This research has made use of the SIMBAD database,
operated at CDS, Strasbourg, France.
This research made use of Montage, funded by the National Aeronautics and 
Space Administration's Earth Science Technology Office, Computational 
Technologies Project, under Cooperative Agreement Number NCC5-626 between NASA
and the California Institute of Technology. The code is maintained by the 
NASA/IPAC Infrared Science Archive.
IRAF is the Image Analysis and Reduction Facility made
available to the astronomical community by the National Optical
Astronomy Observatories, which are operated by AURA, Inc., under
contract with the U.S. National Science Foundation. STSDAS is
distributed by the Space Telescope Science Institute, which is
operated by the Association of Universities for Research in Astronomy
(AURA), Inc., under NASA contract NAS 5--26555.
 Funding for the Sloan Digital Sky Survey (SDSS) and SDSS-II has been provided by the 
 Alfred P. Sloan Foundation, the Participating Institutions, the National Science Foundation, 
 the U.S. Department of Energy, the National Aeronautics and Space Administration, 
 the Japanese Monbukagakusho, and 
 the Max Planck Society, and the Higher Education Funding Council for England. 
 The SDSS Web site is http://www.sdss.org/.
 The SDSS is managed by the Astrophysical Research Consortium (ARC) for the Participating Institutions. 
 The Participating Institutions are the American Museum of Natural History, Astrophysical Institute Potsdam, 
 University of Basel, University of Cambridge, Case Western Reserve University, The University of Chicago, 
 Drexel University, Fermilab, the Institute for Advanced Study, the Japan Participation Group, 
 The Johns Hopkins University, the Joint Institute for Nuclear Astrophysics, the Kavli Institute for 
 Particle Astrophysics and Cosmology, the Korean Scientist Group, the Chinese Academy of Sciences (LAMOST), 
 Los Alamos National Laboratory, the Max-Planck-Institute for Astronomy (MPIA), the Max-Planck-Institute 
 for Astrophysics (MPA), New Mexico State University, Ohio State University, University of Pittsburgh, 
 University of Portsmouth, Princeton University, the United States Naval Observatory, and the University 
 of Washington.

\end{acknowledgements}

\end{document}